\documentclass[usenatbib]{mn2e}
\usepackage{cite}
\usepackage{graphicx}
\usepackage{epsfig}
\usepackage{amsmath,amssymb}
\usepackage{lscape}
\usepackage{verbatim}
\usepackage{rotating}
\usepackage{epstopdf}
\usepackage{rotating}
\usepackage{pdflscape}
\usepackage{subfigure}
\usepackage{chngcntr}
\usepackage{float}
\usepackage{deluxetable}
\usepackage{placeins}
\usepackage{color}

\def\lapprox{\lower.4ex\hbox{$\;\buildrel <\over{\scriptstyle\sim}\;$}}
\def\gapprox{\lower.4ex\hbox{$\;\buildrel >\over{\scriptstyle\sim}\;$}}

\title[Unifying X-ray Scaling Relations]{Unifying X-ray Scaling Relations from Galaxies to Clusters}
\author[Anderson et al.]{Michael E. Anderson$^{1}$\thanks{email: michevan@mpa-garching.mpg.de}, Massimo Gaspari$^{1}$, Simon D. M. White$^{1}$,  Wenting Wang$^{2}$,\newauthor Xinyu Dai$^{3}$ \\
$^{1}$Max-Planck Institute for Astrophysics, Garching bei Muenchen, Germany\\
$^{2}$Institute for Computational Cosmology, Department of Physics, University of Durham, South Road, Durham, DH1 3LE, UK\\
$^{3}$Homer L. Dodge Department of Physics and Astronomy, University of Oklahoma, Norman, OK 73019, USA}

\begin{document}

\maketitle

\begin{abstract}

We examine a sample of $\sim 250 000$ ``locally brightest galaxies'' selected from the Sloan Digital Sky Survey to be central galaxies within their dark matter halos. We stack the X-ray emission from these halos, as a function of the stellar mass of the central galaxy, using data from the ROSAT All-Sky Survey. We detect emission across almost our entire sample, including emission which we attribute to hot gas around galaxies spanning a range of 1.2 dex in stellar mass (corresponding to two nearly orders of magnitude in halo mass) down to $M_{\ast} = 10^{10.8} M_{\odot}$ ($M_{500} \approx 10^{12.6} M_{\odot}$). Over this range, the X-ray luminosity can be fit by a power-law, either of stellar mass or of halo mass. From this, we infer a single unified scaling relation between mass and $L_X$ which applies for galaxies, groups, and clusters. This relation has a slope steeper than expected for self-similarity, showing the importance of non-gravitational heating. Assuming this non-gravitational heating is predominately due to AGN feedback, the lack of a break in the relation shows that AGN feedback is tightly self-regulated and fairly gentle, in agreement with the predictions of recent high-resolution simulations. Our relation is consistent with established measurements of the $L_X$-$L_K$ relation for elliptical galaxies as well as the $L_X$-$M_{500}$ relation for optically-selected galaxy clusters. However, our $L_X$-$M_{500}$ relation has a normalization more than a factor of two below most previous relations based on X-ray-selected cluster samples. We argue that optical selection offers a less biased view of the $L_X$-$M_{500}$ relation for mass-selected clusters.

\end{abstract}
 

\section{Introduction}

Galaxy cluster scaling relations are fundamental tools for connecting cluster astrophysics to cosmology. These scaling relations are typically expressed within the framework of the self-similar model \citep{Kaiser1986}, which predicts that power-law relations should exist between basic properties of clusters (mass, luminosity, temperature, etc.) unless some physical process occurs to produce a characteristic scale and break the self-similarity. 

In this work the scaling relation of particular interest is the $L_X$-$M$ relation. This relation connects the X-ray luminosity of the hot gas in a galaxy cluster (ideally the bolometric luminosity) to the total mass $M$ of the cluster. The self-similar prediction for this relation is that $L_X\propto M^{4/3}$ \citep{Sarazin1986} assuming the X-ray luminosity is dominated by thermal bremsstrahlung. 

A number of studies have estimated the $L_X$-$M$ relation for galaxy clusters (e.g. \citealt{Stanek2006}, \citealt{Maughan2007}, \citealt{Rykoff2008} \citealt{Vikhlinin2009}, \citealt{Pratt2009}, \citealt{Mantz2010}, \citealt{Planck2011}, \citealt{Wang2014}). They typically find slopes in the range of 1.6-2.0, which is significantly steeper than the predicted value of 4/3. Some of this discrepancy may be due to redshift evolution and {mass-dependence of the  temperature and density profiles (\citealt{Vikhlinin2009}, \citealt{Kravtsov2012}), but the majority is usually attributed to non-gravitational heating, either by changing the entropy of the gas \citep{Evrard1990} or through its indirect impact on the baryon fraction of the halo \citep{Mushotzky1997}. 

Non-gravitational heating is also thought to play an important role in the formation and evolution of galaxies and galaxy groups, and there are a number of theoretical predictions for the behavior of the X-ray luminosity of these lower-mass systems.  Generally these studies focus on the $L_X$-$T$ relation rather than the $L_X$-$M$ relation, but temperature and mass are tightly linked in hydrostatic equilibrium, so these relations have similar qualitative behavior. A common prediction from large-scale cosmological simulations (e.g. \citealt{Sijacki2007}, \citealt{Puchwein2008}, \citealt{Fabjan2010}, \citealt{McCarthy2010}, but also see \citealt{LeBrun2014}) is a steepening of the decline in $L_X$ as the halo mass decreases, below $T\sim 1$ keV ($M_{500} \sim 10^{13.5} M_{\odot}$). In many cases, this break is caused by active galactic nucleus (AGN) feedback. Specifically, as pointed out by \citet{Planelles2014} and \citet{Gaspari2014}, the so-called ``thermal blast'' prescription for AGN feedback in these simulations raises the cooling time of the intragroup gas above the Hubble time, converting these galaxy groups into non-cool-core objects. In contrast, gentler ``self-regulated'' mechanical feedback (acting through outflows which induce X-ray buoyant bubbles, weak shocks, and the uplift of low-entropy gas and metals) preserves the cool core and therefore  typically produces no break\footnote{The  terms ``thermal blast'' and ``self-regulated'' are intended to convey the qualitative difference between these types of feedback, although it should be noted that powerful thermal blasts also provide a form of self-regulation. Roughly the same sense is conveyed by substituting the terms ``violent'' and ``gentle'' for the respective types of feedback.} (\citealt{Gaspari2011}, \citealt{Gaspari2012}). 

We can therefore learn more about AGN feedback if we can extend observations of cluster X-ray scaling relations down to the regime of galaxies and galaxy groups. This has posed a formidable challenge, however, due to the lower X-ray luminosity of these less massive systems. Previous observational studies of galaxy groups disagree about whether a $\sim 1$ keV break exists in the $L_X$-$T$ relation (\citealt{Ponman1996}, \citealt{Helsdon2000b}, \citealt{Mulchaey2000}, \citealt{Osmond2004}, \citealt{Sun2009}), and the $L_X$-$M$ relation has only recently begun to be explored at these scales (e.g. \citealt{Bharadwaj2014}, \citealt{Lovisari2014}). So far, systematic studies of the $L_X$-$M$ relation in low-mass groups and isolated galaxies have proven beyond the reach of current X-ray telescopes.

Other X-ray properties and scaling relations have been studied in galaxy-mass halos, however. The closest analogue to the $L_X$-$M$ relation is probably the $L_X$-$L_B$ relation (or for more modern observations, the $L_X$-$L_K$ relation) in elliptical galaxies, which relates the stellar content of an elliptical galaxy to the X-ray properties of its hot gaseous halo. The slope of this relation is also interesting, since it gives clues about the processes which govern the hot gaseous halo.

Unfortunately, the slope of this relation is difficult to measure either as a function of $L_B$ or $L_K$. It seems to depend somewhat on the environment of the galaxies \citep{Mulchaey2010} as well as on the degree of rotational support of the galaxy \citep{Sarzi2013}. Even worse, the scatter in both $L_X$-$L_B$ and $L_X$-$L_K$ is extremely large (an order of magnitude at $10^{11} L_{\odot}$; \citealt{Boroson2011}) which makes sample selection a particularly difficult issue. Sample selection is also a major issue for the $L_X$-$M$ relation, since nearly every study (with the notable exceptions of \citealt{Dai2007}, \citealt{Rykoff2008}, and \citealt{Wang2014}) relies on X-ray-selected clusters. This raises the issue of Malmquist bias\footnote{Malmquist bias is the tendency to overestimate the average luminosity of a population when observing a flux-limited sample, since the more luminous objects are easier to detect.}, which can have a very significant effect on the inferred $L_X$-$M$ relation \citep{Stanek2006}. 

In this study, we take a significant step towards alleviating all of the above issues. We examine an optically-selected sample of central galaxies, which is sensitive to different selection effects than X-ray selected samples. The sample contains $\sim 250 000$ galaxies, each selected to be the most luminous galaxy in its dark matter halo, spanning halos from intermediate-mass clusters down to galaxies about half the mass of the Milky Way. The  selection criteria and basic properties of our sample are described in Section 2. 

With our sample of uniformly-selected central galaxies, we can make the a uniform comparison of the X-ray luminosities of halos across galaxies, galaxy groups, and galaxy clusters for the first time. In order to make this comparison most effectively, and to detect the X-ray emission in lower-mass halos, we employ a stacking technique. This technique is detailed in Section 3. We present the results of our stacking in Section 4. Finally, in Section 5 we measure the $L_X$-$M_{500}$ relation and in Section 6 we measure the $L_X$-$M_{\ast}$ relation. For both relations, a single power-law describes the data from galaxy clusters all the way down to Milky-Way-mass halos. This suggests that the two relations are actually the same. We discuss the implications of this result in Section 7.

\section{Sample}

In this paper we examine the sample of ``locally brightest galaxies'' (LBGs) introduced in \citet{Planck2013} (hereafter P13). Full details of the sample selection are presented in that work, but we briefly summarize them here. 

The LBGs are chosen from the New York University Value-Added Galaxy Catalog  \citep{Blanton2005} based on Data Release 7 from the Sloan Digital Sky Survey (SDSS). The goal is to select a population of central galaxies in dark matter halos. To do this, P13 selected galaxies with extinction-corrected Petrosian $r$-magnitude $r < 17.7$ and redshift $z > 0.03$ which are brighter in $r$ than any other galaxy within 1 projected Mpc and 1000 km s$^{-1}$. To account for potential satellites without spectroscopic redshifts, locally brightest galaxies were also required not to have any galaxies in the SDSS photometric redshift 2 catalog \citep{Cunha2009} with brighter $r$-magnitudes projected within 1 Mpc and having a photometric redshift with a greater than 10\% chance of being consistent with the redshift of the locally brightest galaxy. The total sample contains 259567 LBGs.

For each LBG, stellar masses have been estimated from the SDSS photometry by \citet{Blanton2007}. Following P13, we divided the sample into 20 bins, logarithmically spaced in stellar mass from log $(M_{\ast}/M_{\odot}) = 10.0$ to log $(M_{\ast}/M_{\odot}) = 12.0$ (there are 9660 LBGs which lie outside this range in stellar mass and are discarded; 35 are above the upper limit and 9625 are below the lower limit). The distribution of galaxies across stellar mass is presented in Figure 1. The sample is approximately flux-limited: SDSS photometric completeness for galaxies at $r = 17.7$ is essentially 100\%, although spectroscopic completeness is lower due to fiber ``collisions'' where two or more galaxies fall within the same SDSS fiber. Checking the \verb"FGOT" parameter from the Value-Added Galaxy Catalog, we find approximately 91\% completeness for the locally brightest galaxy sample. We also note that the selection criteria may introduce subtle and complex biases compared to a purely flux-limited sample.

\begin{figure}
\begin{center}
\includegraphics[width=8.5cm]{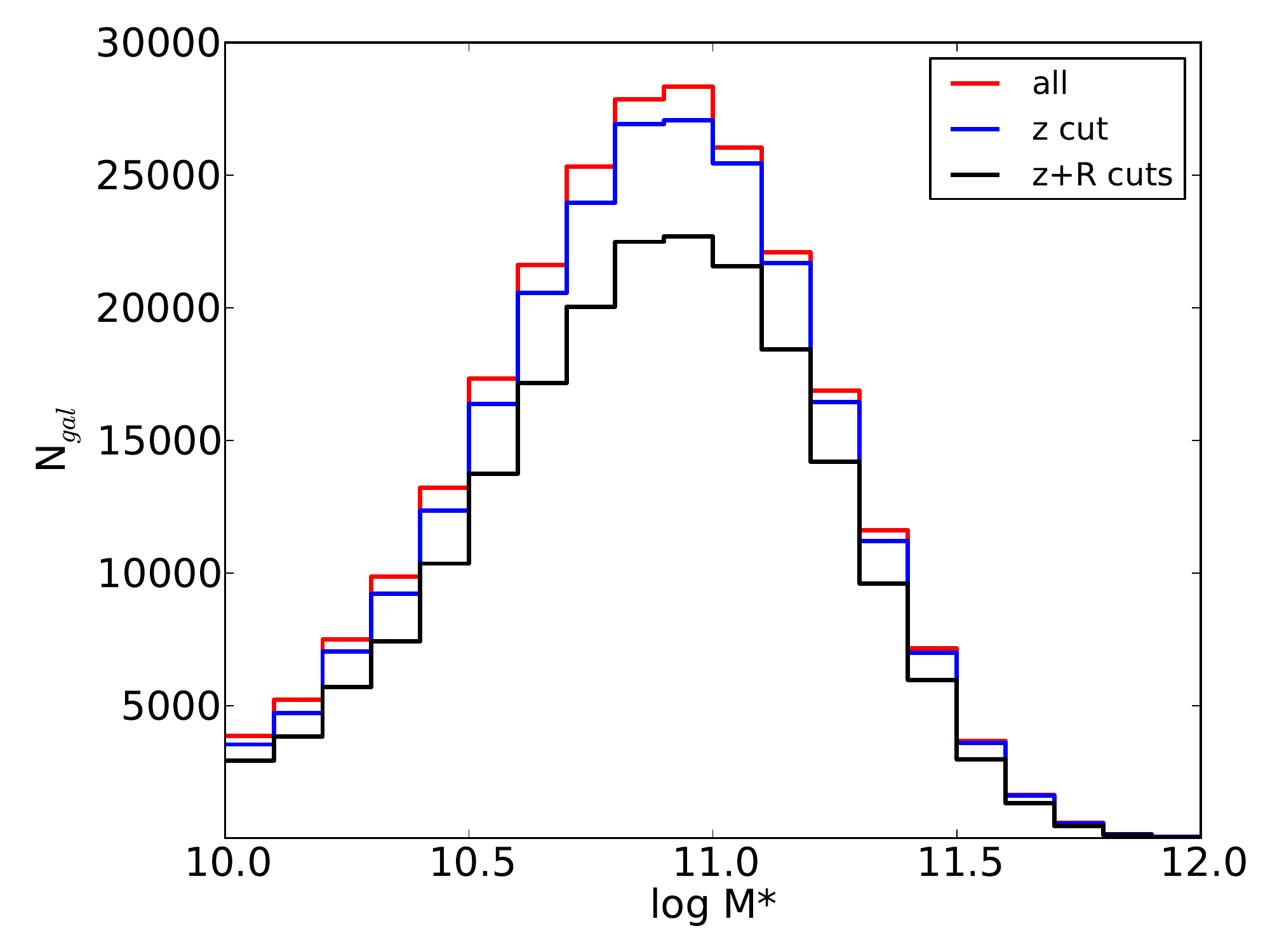}
\end{center}
\caption{Stellar mass distribution of locally brightest galaxies. The red line shows all 249907 LBGs within our stellar mass limits. The blue line shows the 239389 LBGs within our redshift limits. The black line shows the 201011 LBGs that fall within our redshift limits and do not overlap with the edge of one of the slices in the RASS. }
\end{figure}

In order to better understand these biases, we employ a catalog of simulated locally brightest galaxies originally generated for P13. The procedure for generating this catalog starts with the Millennium simulation \citep{Springel2005} which follows the evolution of cosmic structure within a box of comoving side length 500 $h^{-1}$Mpc. Halo merger trees are complete for subhalos above a mass of $1.7\times10^{10} h^{-1}\mathrm{M_\odot}$. The original Millennium simulation is based on WMAP1 cosmology, and the rescaling technique of \citet{Angulo2010} has been adopted to convert it to the WMAP7 cosmology. A semi-analytic galaxy formation prescription \citep{Guo2011} has been applied to these simulated halos. The galaxy formation parameters have been adjusted to fit several statistical observables at $z=0$, such as the luminosity, stellar mass and correlation functions of galaxies. We project the simulation box along the $z$-axis, and assign every galaxy an artificial redshift based on its line-of-sight distance and peculiar velocity, i.e., parallel to the $z$-axis. In this way we can select a sample of galaxies from the simulation using isolation criteria exactly analogous to those used for the locally brightest galaxy sample from SDSS. 

These simulations allowed P13 to match the stellar content of galaxies to dark matter halos from the Millennium Simulation, as well as to estimate the fraction of satellite galaxies or other failure modes in the sample. They find that the fraction of locally brightest galaxies which are centrals is over 83\% across the sample. About 2/3 of the satellite LBGs are brighter than the central galaxy, while the other 1/3 are either more than 1 Mpc from the central or are offset in velocity by more than 1000 km s$^{-1}$. They also are able to estimate the relationship between $M_{\ast}$ and halo mass for galaxies in each stellar mass bin. They find that the satellite contamination biases the mean halo mass upwards fairly significantly in the lower-mass bins, but in general these satellite LBGs are significantly offset from the center of their massive halos and therefore do not appreciably bias the stacked signal. We therefore treat the terms  ``locally brightest galaxy'' and ``central galaxy'' as approximately interchangeable throughout this work.

This simulated catalog also allows estimation of ``effective'' halo masses for each stellar mass bin down to log $M_{\ast} = 10.8-10.9$. P13 computed halo masses by assuming self-similar scaling in the $Y_{SZ}$-$M_{500}$ relation and assuming the \citet{Arnaud2010} pressure profile for the hot gas. Since X-ray luminosity scales with the projected squared density instead of the projected density, the effective halo masses are slightly higher for X-ray emission. Also, one of our conclusions is that the $L_X$-$M$ relation is not self-similar, so we instead use our best-fitting relation when computing the effective halo masses. In Appendix A, we detail the derivation of our effective halo masses and compare with the masses used in P13; our effective halo masses are similar to those in P13, and the halo masses seem to be fairly robust.

The stellar masses, halo masses, and other derived quantities in both P13 and this work rely on an assumed WMAP7 cosmology \citep{Komatsu2011}, with $\Omega_m = 0.272$, $\Omega_{\Lambda} = 0.728$, and $H_0 = 70.4$ km s$^{-1}$ Mpc$^{-1}$. We generally work with halo masses expressed as $M_{500}$, the mass within a radius $R_{500}$ which encloses a mean density 500 times the critical density. We estimate $R_{500}$ from $M_{500}$ using the relation

\begin{equation}
M_{500} \equiv 500 \times \frac{4}{3} \pi R_{500}^3 \rho_c
\end{equation}

The critical density of the Universe, $\rho_c$, has a redshift dependence, and we compute it at the mean distance $\overline{d_L}$ of the galaxies in the bin, which is defined in section 3.1. This redshift effect is small, since the highest effective redshift of any bin is 0.29. For each stellar mass bin, the effective values of $M_{500}$ and $R_{500}$ we adopt are listed below in Table 1, along with $\overline{d_L}$ and several other properties.

\begin{table*}
\begin{minipage}{120mm}
\caption{Parameters for Locally Brightest Galaxies in Stellar Mass Bins}
\begin{tabular}{ccccccccc}
\hline
log $M_{\ast}$ & log $M_{500}$ & $R_{500}$ & $R_{\text{extract}}$ & $z_{\text{min}}$ & $z_{\text{max}}$ & $N_{\text{gal}}$ & $N_{\text{stacked}}$ & $\overline{d_L}$\\
($M_{\odot}$) & ($M_{\odot}$) & (kpc) & (kpc) & & & & & (Mpc)\\
\hline
11.9-12.0 & 14.56 & 938 & 3000 & 0.15 & 0.40 & 44 & 36 & 1492.7  \\
11.8-11.9 & 14.41 & 838 & 3000 & 0.15 & 0.40 & 145 & 114 & 1433.7  \\
11.7-11.8 & 14.29 & 772 & 2500 & 0.10 & 0.35 & 573 & 455 & 1275.7  \\
11.6-11.7 & 14.08 & 665 & 2000 & 0.10 & 0.35 & 1624 & 1326 & 1088.5\\
11.5-11.6 & 13.90 & 584 & 2000 & 0.10 & 0.30 & 3664 & 2967 & 1009.8\\
11.4-11.5 & 13.70 & 504 & 1500 & 0.10 & 0.30 & 7160 & 5970 & 915.6  \\
11.3-11.4 & 13.51 & 437 & 1000 & 0.07 & 0.25 & 11615 & 9615 & 788.6 \\
11.2-11.3 & 13.29 & 372 & 1000 & 0.07 & 0.25 & 16871 & 14194 & 714.6 \\
11.1-11.2 & 13.09 & 320 & 1000 & 0.06 & 0.25 & 22085 &18430 & 633.9  \\
11.0-11.1 & 12.91 & 280 & 1000 & 0.06 & 0.25 & 26026 & 21583 & 592.0 \\
10.9-11.0 & 12.75 & 248 & 1000 & 0.05 & 0.20 & 28325 & 22689 & 523.1 \\
10.8-10.9 & 12.60 & 222 & 1000 & 0.05 & 0.20 & 27866 & 22490 & 485.3 \\
10.7-10.8 & 12.34 & 182 & 1000 & 0.05 & 0.18 & 25309 & 20041 & 455.4 \\
10.6-10.7 & 12.20 & 164 & 1000 & 0.05 & 0.18 & 21619 & 17168 & 428.2 \\
10.5-10.6 & 12.09 & 150 & 1000 & 0.05 & 0.18 & 17328 & 13729 & 407.0 \\
10.4-10.5 & 11.99 & 140 & 1000 & 0.05 & 0.18 & 13221 & 10353 & 386.3 \\
10.3-10.4 & 11.90 & 131 & 1000 & 0.04 & 0.14 & 9862 & 7425 & 339.1 \\
10.2-10.3 & 11.82 & 124 & 1000 & 0.04 & 0.14 & 7499 & 5693 & 325.1  \\
10.1-10.2 & 11.75 & 117 & 1000 & 0.04 & 0.12 & 5223 & 3821 & 308.0 \\
10.0-10.1 & 11.69 & 111 & 1000 & 0.04 & 0.12 & 3848 & 2912 & 298.6 \\
\hline
\end{tabular}
\\
\small{Properties of LBGs in each of our 20 stellar mass bins. Stellar masses are measured from SDSS photometry, halo masses are estimated from simulations as described in Appendix A, and $R_{500}$ is estimated from the halo mass using equation 1.  $R_{\text{extract}}$, $z_{\text{min}}$, and $z_{\text{max}}$ are defined in section 3.1. $N_{\text{gal}}$ is the total number of LBGs in the bin, and $N_{\text{stacked}}$ is the number which pass our additional selection criteria in section 3.1 and are included in the stacks. The mean distance of the stacked galaxies, $\overline{d_L}$, is described in section 3.1 as well.  }
\end{minipage}
\end{table*}

\section{Stacking}

We examine stacked images of these galaxies using data from the ROSAT All-Sky Survey (RASS). The dataset is a shallow all-sky survey in the soft X-rays. The stacking software was originally generated by \citet{Dai2007}, and the stacking procedure here is very similar to that used in \citet{Anderson2013} (hereafter ABD13). In brief, for each of the 20 samples of LBGs binned by stellar mass (Table 1), we follow this procedure:

1. Extract a RASS image of each LBG and its surroundings, extending in physical space out to $R_{\text{extract}}$. Also extract the corresponding RASS exposure map for each field.

2. Exclude bright point sources from each image, using a combination of cross-matching with the ROSAT Bright Source Catalog and Faint Source Catalog \citep{Voges1999} and automatic flagging of pixels with large count rates. 

3. Add together each of the RASS images to produce a stacked image in physical space. Also add together each of the exposure maps, weighting the individual exposure maps by a factor proportional to the angular area covered by the field.

4. Construct an empirical point spread function (psf) for the stacked image, by stacking at least 10000 known point sources from the ROSAT Faint Source Catalog and ROSAT Bright Source Catalog using the same apertures and weighting scheme as the LBGs.

5. Perform aperture photometry on the stacked image in order to estimate average background-subtracted count rates. We use two apertures: a circle with radius corresponding to the average $R_{500}$ for the LBGs in the stack, and an annulus extending from $(0.15 - 1)\times R_{500}$.

6. Convert the measured count rates into an average X-ray luminosity for each stack.

Stacking in physical space instead of angular space is a natural choice for this project, since it allows us to study the galactic and circumgalactic emission as a function of radius around locally brightest galaxies. However, this choice does complicate the analysis, in three major ways. First, it means the aperture will vary in angular size from galaxy to galaxy, which will naturally over-weight the nearest galaxies, since their apertures will be the largest and therefore enclose the most photons. Second, the point spread function becomes more complicated, since the psf is typically defined in angular units. Finally, the varying apertures also introduce complications for the nearest and the most distant galaxies in each stellar mass bin. In the rest of this section, we describe our procedure in more detail, including how we account for the complications listed above.

\subsection{Additional Details}

The first issue in our procedure, similar to ABD13, is selection of minimum and maximum redshifts $z_{\text{min}}$ and $z_{\text{max}}$ for locally brightest galaxies in each bin. We choose these redshifts to enclose as much of the sample as possible, with the additional requirement that $z_{\text{min}}$ to be about a third of $z_{\text{max}}$. This excludes the handful of galaxies with the lowest redshifts, preventing them from dominating the overall signal due to the larger apertures necessary to enclose these galaxies. A handful of the most distant galaxies are also excluded, although their apertures are so small that they add comparatively little signal. The redshift distributions of the LBG samples and the locations of $z_{\text{min}}$ and $z_{\text{max}}$ can be seen in Figure 2.

\begin{figure}
\begin{center}
\includegraphics[width=8.5cm]{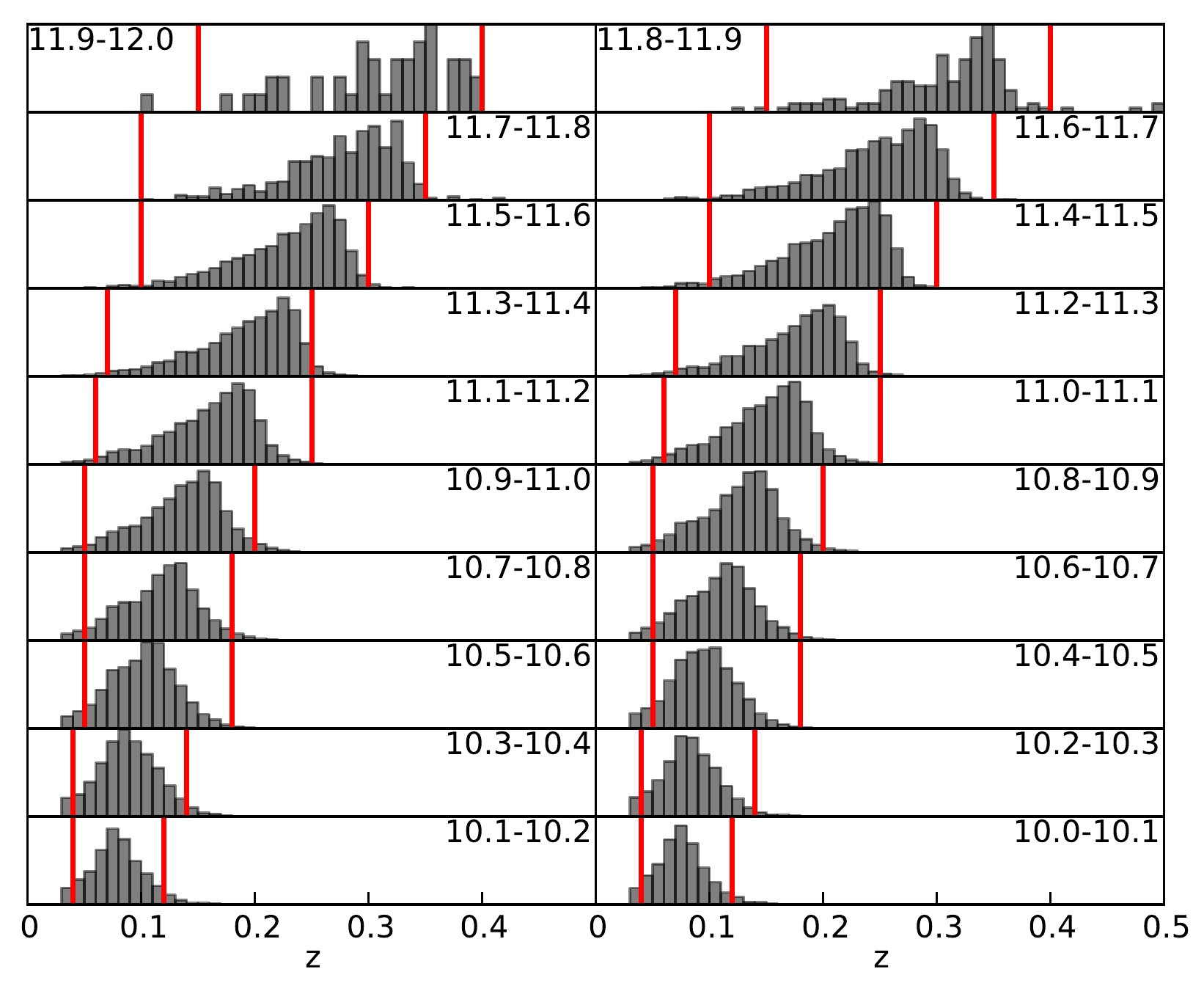}
\end{center}
\caption{Redshift distributions of locally brightest galaxies in each of our 20 stellar mass bins. In each panel, the range of log $M_{\ast}$ is indicated, and the redshift cuts are denoted by the vertical red lines. The total number of LBGs in each bin varies, as shown in Figure 1.}
\end{figure}

The next issue is the selection of $R_{\text{extract}}$. The natural choice for  $R_{\text{extract}}$ is 1000 kpc, since this is the projected radius within which the isolation criterion is defined. We choose this radius for most stellar mass bins, but the largest LBGs have values of $R_{500}$ close to 1000 kpc, so we extend $R_{\text{extract}}$  to larger values for these systems. As $R_{\text{extract}}$ increases, we become increasingly  limited by the size of the individual fields ($6.4^{\circ} \times 6.4^{\circ}$) which comprise the RASS. In ABD13 we were able to model the so-called ''effective vignetting'' introduced when the extraction region extended beyond the edge of one of these fields, but for our simpler aperture photometry in this work, we exclude any galaxy from our sample if the $R_{\text{extract}}$ aperture overlaps with the nearest edge of its RASS field. The effect of this exclusion can be seen in the blue histogram in Figure 1. 

With $z_{\text{min}}$, $z_{\text{max}}$, and $R_{\text{extract}}$ defined, we can now extract RASS images and exposure maps for each LBG. We extract images in the 0.5-2.0 keV band, and compute  the angular size of $R_{\text{extract}}$ for each galaxy using its SDSS redshift and assuming WMAP7 cosmology. We create images which are $200\times200$ pixels in size. This means that, for the most distant galaxies in each bin, the effective pixel size is often a bit smaller than the ROSAT $1\sigma$ pointing accuracy of 6''. In practice this is not a major concern: the most distant galaxies contribute the fewest photons, our empirical psf technique should account for this effect where it exists, and we analyze the images using aperture photometry with large apertures so that single-pixel accuracy is not particularly important. 

The next issue is how to treat bright point sources that randomly lie within the images. We use a similar approach as ABD13, masking out any portion of an image with a source listed in the ROSAT Bright Source catalog (BSC) or the ROSAT Faint Source catalog (FSC). Many locally brightest galaxies are listed in these catalogs, however, so for each bin we impose a minimum count rate (using the count rates for these sources listed in the catalogs) for sources to be excluded. These count rates are tabulated in Appendix B, and correspond to luminosities (if the source were at the mean distance of the LBGs in each bin) which are more than an order of magnitude above the mean LBG luminosity. Additionally, we exclude any observation that has more than 10 counts in a single pixel, although this too has little effect on the results. We explore the effect of changing the minimum count rate and removing the 10 count threshold in Appendix B, and find that these changes have little effect on the results.  

To generate the stacked images, we add the individual images without any weighting. Thus the stacked images contain integer numbers of photons and are subject to the usual Poisson statistics. We generate a composite exposure map as well, by stacking the individual exposure maps, but we weight the exposure maps in the stack in order to account for the differences in aperture sizes due to the different distances of the LBGs. In general, locally brightest galaxies are either undetected or marginally detected in the RASS, so the images are background-dominated and the appropriate weighting is therefore proportional to the area subtended by the image\footnote{This assumes a spatially uniform X-ray background, which is an acceptable approximation for this analysis since any variations in the background are averaged out over the large number of images in each stack.}. In Figure 3, we present the stacked images in each of our 20 stellar mass bins.

\begin{figure*}
\begin{center}
\includegraphics[width=15.5cm]{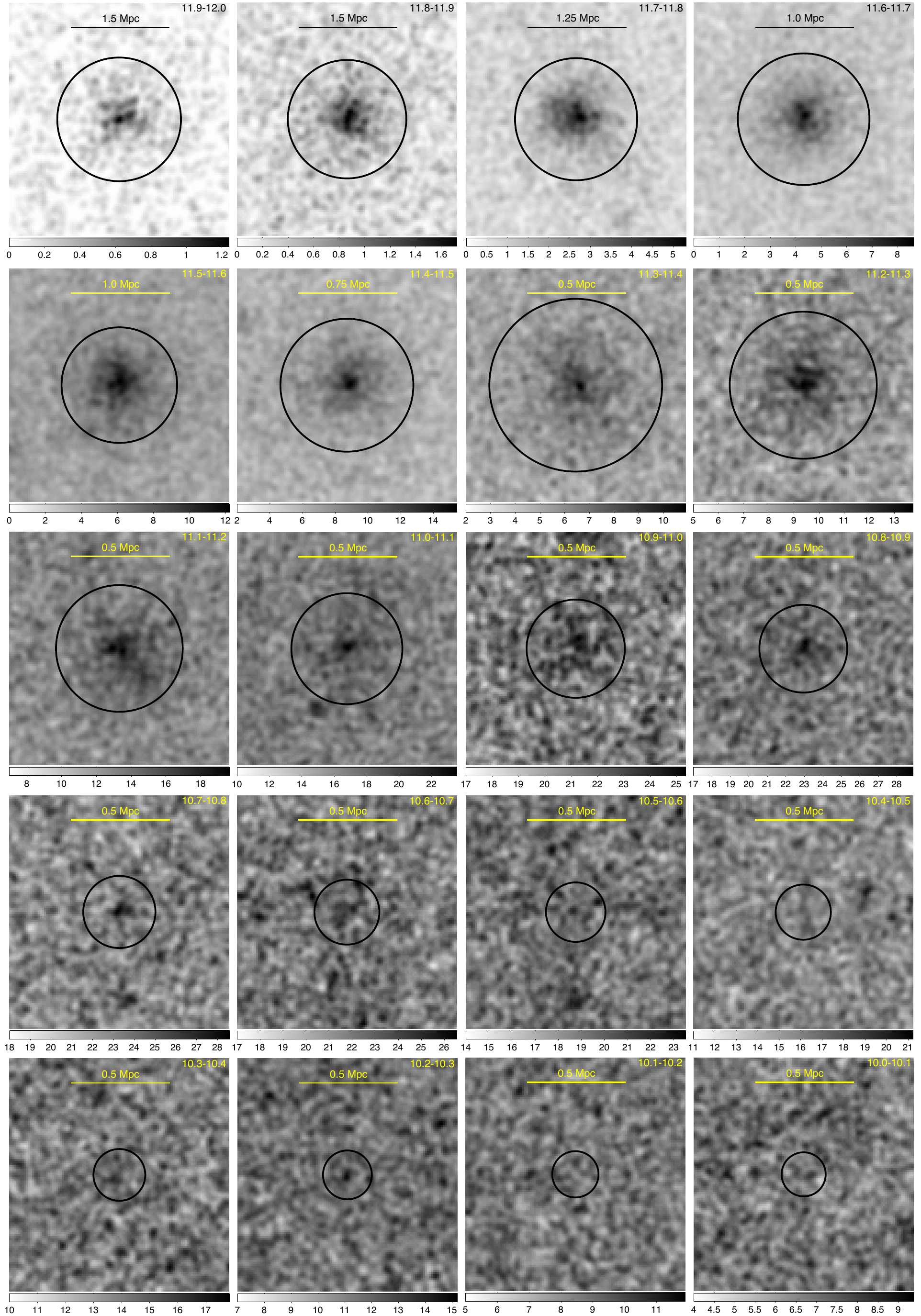}
\end{center}
\caption{Stacked 0.5-2.0 keV RASS images of locally brightest galaxies in each of our 20 stellar mass bins. In each image, log $M_{\ast}$ is noted at the top right and $R_{500}$ is indicated with the black circle. Note that the physical scale and colorbar vary across these images. Each image has been smoothed with a Gaussian 3-pixel kernel.}
\end{figure*}

Next, we construct empirical psfs to match each stacked image. To do this, we stack at least $10^4$ point sources from the combination of the ROSAT Faint Source Catalog and the ROSAT Bright Source Catalog \citep{Voges1999} matched in count rate distribution to the expected count rate distribution of the galaxies in each stack, and we assign apertures to each source matching the aperture distribution used for the corresponding stack as well. For more details on this process, see ABD13. These psfs are used as part of the aperture photometry, as described below.

In Figure 4, we present radial surface brightness profiles for each of our bins, centered on the locally brightest galaxy. The empirical psf is also indicated, normalized to the count rate in the center of the image. We also indicate the fit to the background with the horizontal line and the adopted value of $R_{500}$ with the vertical line in each image.

\begin{figure*}
\begin{center}
\includegraphics[width=17cm]{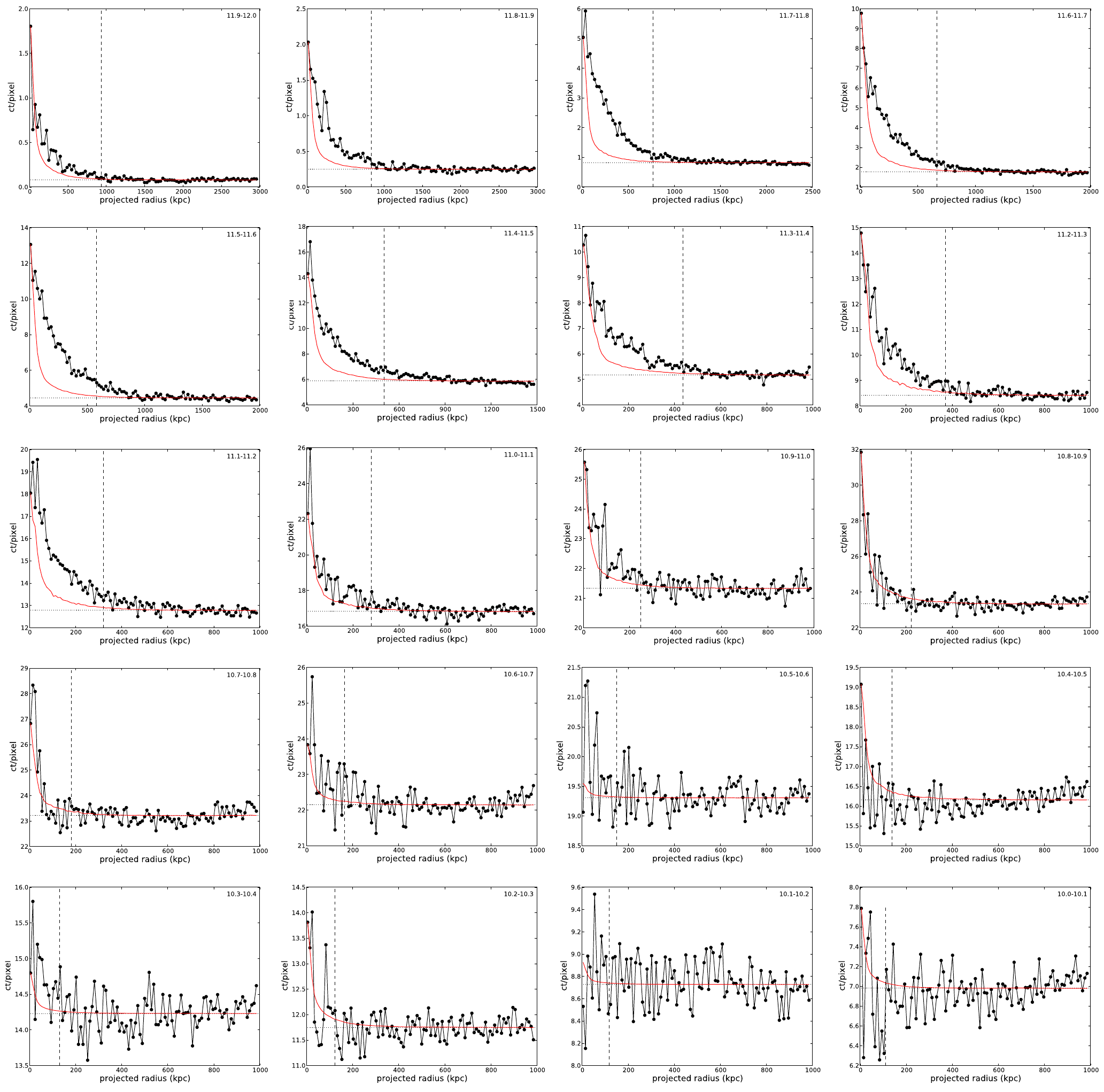}
\end{center}
\caption{Azimuthally averaged surface brightness profiles of the stacked images in Figure 3. In each image, log $M_{\ast}$ is noted at the top right and $R_{500}$ is indicated with the black dashed vertical line. The black dashed horizontal line indicates the fit to the background. The red line is the shape of the empirical psf in each figure, normalized to match the value of the central bin, and is shown here for illustration of the extended nature of many of these profiles. }
\end{figure*}

We analyze these images using aperture photometry. We specify two different source apertures: the ''total'' aperture is a circle extending out to $R_{500}$ and the ''circumgalactic'' or ''CGM'' aperture is an annulus extending from $(0.15-1)\times R_{500}$. The background is determined from an annulus extending from $1.5\times R_{500}$ to $R_{\text{extract}}$. As Figure 4 shows, the psf is much more compact than $R_{500}$, so we apply no psf correction to the ''total'' aperture, but psf effects can be important when performing the ''CGM'' photometry. For this region, we compute the count rate within the CGM annulus and subtract from it the count rate within $0.15R_{500}$ scaled by the fraction of the power in the psf which falls within the CGM annulus. 

We also allow for uncertainty in $R_{500}$, which stems primarily from uncertainty in $M_{\text{halo}}$ for each bin. We estimate this uncertainty using a catalog of simulated locally brightest galaxies in dark matter halos, described in more detail in section 4. For the eight lowest-mass bins, which are not included in the simulated catalog, we use the abundance-matching relation of \citet{Moster2010} to compute $M_{500}$, and we propagate the uncertainties in this relation through into the estimation of $R_{500}$. 

Finally, we estimate the average X-ray luminosity of the LBGs in each bin. We start with the results of the aperture photometry, which are background-subtracted fluxes within two apertures, in the observer-frame 0.5-2.0 keV band, in units of counts$/$second$/$galaxy. We then k-correct this count rate into the rest-frame 0.5-2.0 keV band, using the average (aperture-area-weighted) luminosity distance of the LBGs in each bin in order to estimate the effective redshift. These effective redshifts are small (maximum $z = 0.29$) and so the k-corrections are also small. To compute the k-correction, we estimate the virial temperature of the LBGs in the bin based on the $M_{500}$-$T$ relation in \citet{Sun2009}. We use their relation calibrated across the full sample of galaxy groups and clusters (``Tier 1 + 2 + clusters''). This relation gives an average value for the temperature with the core of the group/cluster excluded, but we neglect the difference between this value and the emission-weighted temperature with the core included. We then estimate the k-correction assuming an APEC model \citep{Smith2001} with that temperature for the X-ray emission\footnote{In the less-massive halos, X-ray binaries will also contribute significantly to the emission, but here the redshifts are so small that the k-correction is already negligible.}. We then convert the count rate into a flux ($f_X$) by multiplying by temperature-dependent conversion factor computed from the APEC model using the WEBPIMMS\footnote{http://heasarc.gsfc.nasa.gov/cgi-bin/Tools/w3pimms/w3pimms.pl} tool. In the bins with log $M_{\ast} < 10.8$, X-ray binaries can be expected to contribute significantly to the emission as well, so we set this conversion factor to $1.2\times10^{-11}$ erg count$^{-1}$ cm$^{-2}$, which we showed in ABD13 is fairly insensitive to the contribution of X-ray binaries to the hot gas emission. Finally, the luminosity is red derived from the flux $f_X$ according to the relation} $L_X = 4\pi \overline{d_L}^2 f_X $. In Table 2 we present all of these correction factors for each observation.

\begin{table}
\caption{Conversion Factors for Computing the Luminosity of Locally Brightest Galaxies}
\begin{tabular}{ccccc}
\hline
log $M_{\ast}$ & $\overline{kT}$ & k & $c_{\text{flux}}$ & $C_{\text{bolo}}$\\
$(M_{\odot})$ & (keV) & & ($10^{-11}$ erg count$^{-1}$ cm$^{-2}$) & \\
\hline
11.9-12.0 & 5.0 & 0.88 &1.2  & 3.2 \\
11.8-11.9 & 4.0 & 0.90 &1.2 & 2.8\\
11.7-11.8 & 3.4 & 0.91 & 1.2 & 2.5\\
11.6-11.7 & 2.5 & 0.95 & 1.1 & 2.1\\
11.5-11.6 & 1.9 & 0.96 & 1.1 & 1.8\\
11.4-11.5 & 1.5 & 0.98 & 1.1 & 1.4\\
11.3-11.4 & 1.1 &1.00 & 1.0 & 1.3\\
11.2-11.3 & 0.8 &1.02 & 1.0 & 1.2\\
11.1-11.2 & 0.6 &1.03 & 0.9 & 1.1 \\
11.0-11.1 & 0.5 & 1.05 &1.0 & 1.1\\
10.9-11.0 & 0.4 &1.06 &1.0 & 1.1\\
10.8-10.9 & 0.3&1.06 & 1.1 & 1.2\\
10.7-10.8 & 0.2& 1.07 &1.2 & 1.2\\
10.6-10.7 & 0.2& 1.07 & 1.2 &--\\
10.5-10.6 & 0.1& 1.06 & 1.2&--\\
10.4-10.5 & 0.1& 1.06 &1.2&--\\
10.3-10.4 & 0.1& 1.06 &1.2&--\\
10.2-10.3 & 0.1& 1.06 & 1.2&--\\
10.1-10.2 & 0.1& 1.05 & 1.2&--\\
10.0-10.1 & 0.1& 1.05 & 1.2&--\\
\hline
\end{tabular}
\\
\small{Conversion factors and derived average hot gas temperatures for locally brightest galaxies. The mean temperature of the hot gas, $\overline{kT}$, is described in section 3.1. The factor $k$ is the estimated k-correction factor to convert the observed 0.5-2.0 keV flux into the rest frame, and the factor $c_{\text{flux}}$ is the estimated (temperature-dependent) conversion factor between counts and erg cm$^{-2}$. The bolometric correction $C_{\text{bolo}}$ is used in sections 4 and 5 to study scaling relations; this analysis is only performed on the uppermost 12 bins so no bolometric correction is ever applied to the lowest 8 bins.   }
\end{table}

In Appendix C, we perform null tests to verify that our stacking procedure has no intrinsic bias, and in Appendix D we stack simulated data in order to verify that we can recover injected $L_X$-$M_{500}$ relations correctly.

\section{Results}

Here we present the results of the stacking procedure detailed in Section 3. In Table 3 and Figure 5, we present the measured luminosities. The power-law in $L_{\text{total}}$ seems unbroken down to stellar masses as low as log $M_{\ast} = 10.7-10.8$. The flattening below this mass is analogous to the flattening observed in P13, and is largely due to the X-ray signal becoming too faint to distinguish from the background. However, unlike with the Sunyaev-Zel'dovich (SZ) effect in P13, there are additional sources of X-ray emission which become important in low-mass galaxies. While bright AGNs are not a major concern (see Appendix B), we do have to account for low-mass X-ray binaries (LMXBs) and high-mass X-ray binaries (HMXBs). We construct simple estimates of the contribution from these sources in Appendix E, and show the results in Figure 5 as well.

\begin{figure*}
\begin{center}
\subfigure[Total]{\includegraphics[width=12cm]{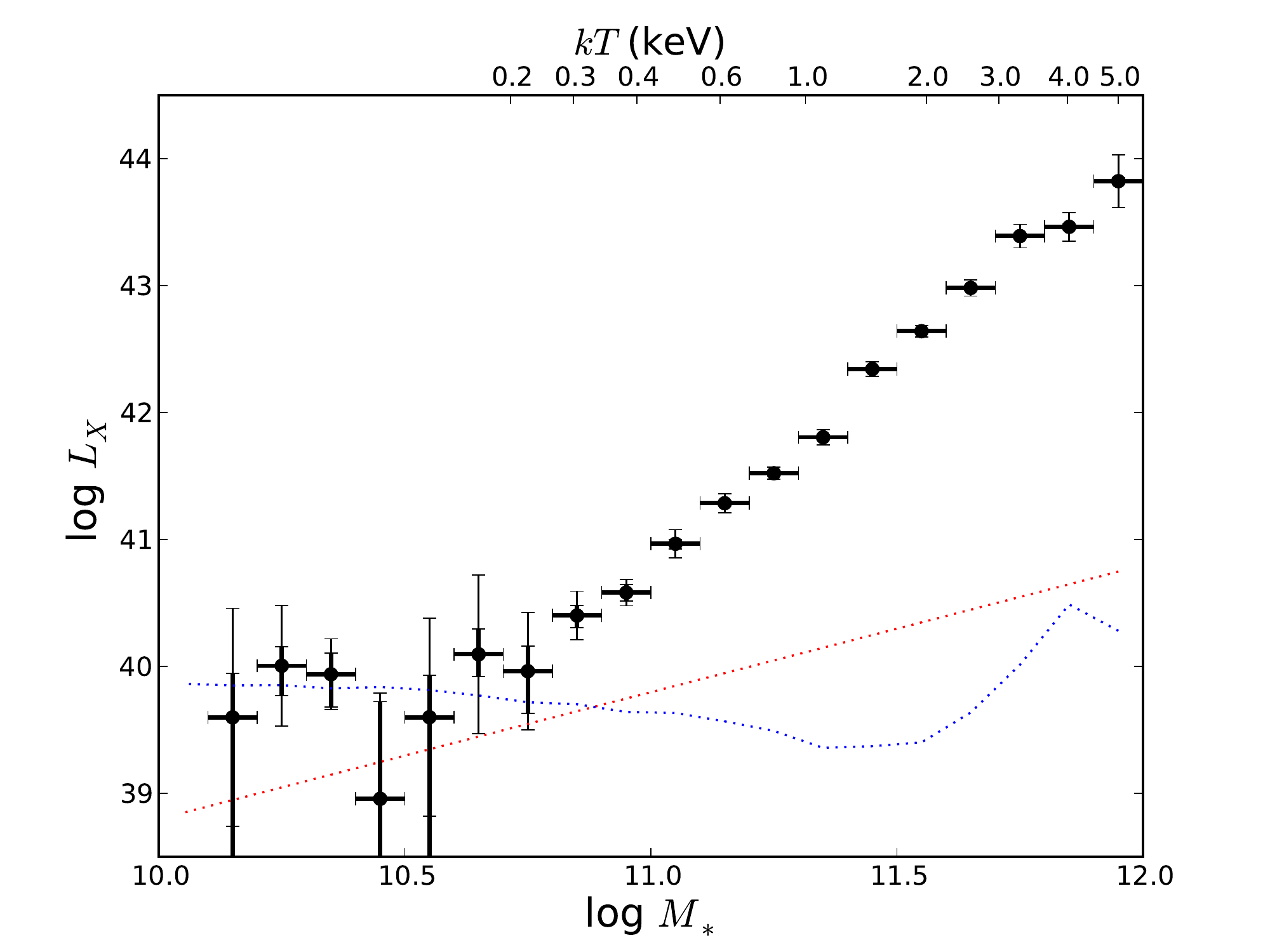}}
\subfigure[CGM]{\includegraphics[width=12cm]{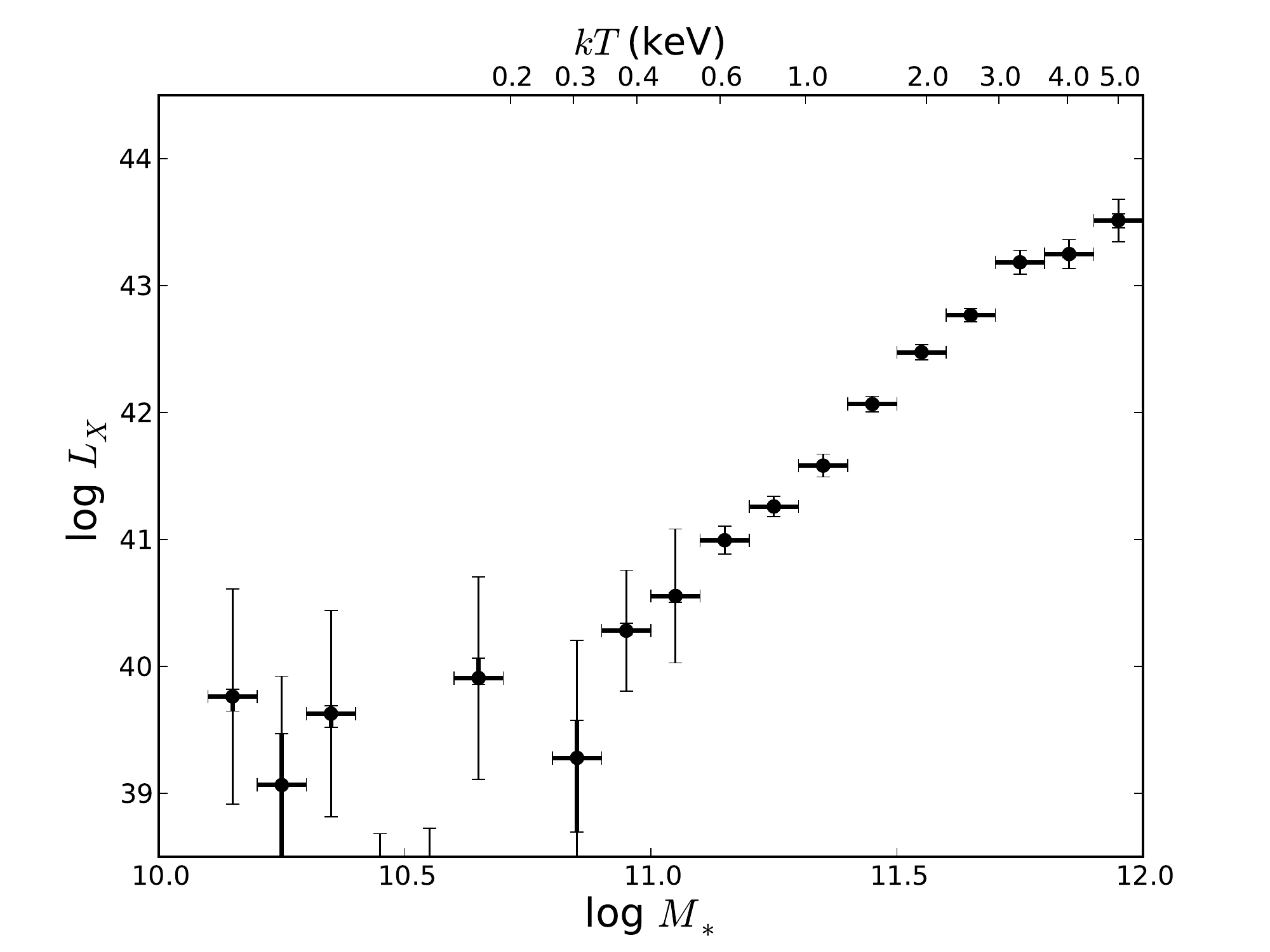}}
\end{center}
\caption{Average 0.5-2.0 keV luminosities of stacked locally brightest galaxies. Panel (a) shows $L_{\text{total}}$, the average luminosity projected within $R_{500}$, and panel (b) shows $L_{\text{CGM}}$, the average luminosity projected between $(0.15-1)\times R_{500}$. In both plots, the thick error bar shows the $1\sigma$ measurement error from photon counting statistics and the thin error bar shows the $1\sigma$ uncertainty in the mean value as determined from bootstrapping analysis. Both $L_{\text{total}}$ and $L_{\text{CGM}}$ obey simple power-law relations with no breaks down to the lowest luminosities where emission can be distinguished from the background. In (a), we also include lines showing the approximate expected contribution from low-mass X-ray binaries (red) and high-mass X-ray binaries (blue) in each bin. In both plots, the upper X-axis axis shows the approximate average gas temperatures, as described in section 3.1.  }
\end{figure*}

\begin{table*}
\begin{minipage}{110mm}
\caption{Luminosities of Locally Brightest Galaxies }
\begin{tabular}{ccccccc}
\hline
log $M_{\ast}$ & log $L_{\text{total}}$ &  $\sigma_m L_{\text{total}}$ & $\sigma_b L_{\text{total}}$ & log $L_{\text{CGM}}$ & $\sigma_m L_{\text{CGM}}$ & $\sigma_b L_{\text{CGM}}$ \\
$(M_{\odot})$ & (erg s$^{-1}$) & (dex) & (dex) & (erg s$^{-1}$) & (dex) & (dex) \\ 
\hline
11.9-12.0 & 43.82 & 0.03 & 0.21 & 43.51 & 0.06 & 0.17 \\
11.8-11.9 & 43.46 & 0.02 & 0.11 & 43.25 & 0.02 & 0.11 \\
11.7-11.8 & 43.39 & 0.01 & 0.09 & 43.18 & 0.01 & 0.09\\
11.6-11.7 & 42.98 & 0.01 & 0.06 & 42.77 & 0.01 & 0.05\\
11.5-11.6 & 42.64 & 0.01 & 0.05 & 42.47 & 0.01 & 0.06\\
11.4-11.5 & 42.34 & 0.01 & 0.06 & 42.07 & 0.02 & 0.06\\
11.3-11.4 & 41.80 & 0.02  & 0.06 & 41.58 & 0.02 & 0.09\\
11.2-11.3 & 41.52 & 0.02 & 0.05 & 41.26 & 0.02 & 0.08 \\
11.1-11.2 & 41.29& 0.02 & 0.07 & 40.99 & 0.02 & 0.11 \\
11.0-11.1 & 40.97 & 0.04 & 0.11 & 40.55 & 0.04 & 0.53\\
10.9-11.0 & 40.58 & 0.07 & 0.10 & 40.28 & 0.05 & 0.48 \\
10.8-10.9 & 40.40 & 0.09 & 0.19 & 39.28 & 0.44 & 0.93\\
10.7-10.8 & 39.96 & 0.27 & 0.46 & $<0$ & -- & 0.30 \\
10.6-10.7 & 40.10 & 0.19 & 0.63 & 39.91 & 0.10 & 0.80 \\
10.5-10.6 & 39.60 & 0.97 & 0.78 & $<0$ & -- & 0.72 \\
10.4-10.5 & 38.96 & 0.86 & 0.83 & $<0$ & -- & 0.68 \\
10.3-10.4 & 39.94 & 0.21 & 0.28 & 39.63 & 0.08 & 0.81\\
10.2-10.3 & 40.00 & 0.19 & 0.47 & 39.07 & 0.73 & 0.86\\
10.1-10.2 & 39.60 & 0.97 & 0.86 & 39.76 & 0.09 & 0.85\\
10.0-10.1 & $<0$ & -- & 0.40 & $<0$ & -- & 0.42 \\
\hline
\end{tabular}
\\
\small{Measured 0.5-2.0 keV average luminosities of locally brightest galaxies. For each luminosity, two $1\sigma$ uncertainties are quoted. The first is the measurement uncertainty (incorporating Poisson uncertainty in the source region, uncertainty in the value of $R_{500}$, and Poisson uncertainty in the level of the background) and the second is the sample error on the mean as estimated from bootstrapping analysis. $1\sigma$ upper limits which are negative are denoted as $<0$. }
\end{minipage}
\end{table*}

Note that essentially all the observed X-ray emission in the lowest 8 bins can be explained as the sum of LMXB and HMXB emission. On the other hand, little of the emission in the highest 12 bins can be explained by X-ray binaries (XRBs). 

The results for $L_{\text{CGM}}$ are consistent with our simple model for XRB emission as well. In seven of the lowest eight bins, the measured luminosity is consistent with or less than zero. This is exactly what we would expect if the signal were dominated by LMXBs and HMXBs from within the galaxy, since the emission from the galaxy is not included in our CGM annulus. The uppermost 11 bins all show secure detections of extended emission which is almost certainly hot gas. In the log $M_{\ast} = 10.8-10.9$ bin, the measured luminosity lies on the same power-law, but the detection of hot extended emission is less secure (see next paragraph). 

We also can estimate the intrinsic variation in $L_X$ within each bin from the sample error on the mean. We do this using bootstrapping. For each bin, we generate 100 bootstrapped samples (with replacement) and stack each sample, applying the same analysis as the real data, and deriving a measurement for $L_X$. The standard deviation of these 100 bootstrapped measurements is plotted for each bin with thin error bars. The sample error on the mean decreases as the number of LBGs in each bin increases, but it still gives some sense of the intrinsic scatter in $L_X$ among LBGs in each bin. The scatter is discussed more in section 5.2 and in Appendix G.

\section{The $L_X$-$M_{500}$ relation}

In this section we derive a simple $L_X$-$M_{500}$ relation from our results for the uppermost 12 bins. As is typical for studies of galaxy clusters, we assume a parametric form for this relation with a power-law dependence on $M_{500}$ and a factor of $E(z)^{7/3}$ to allow for self-similar redshift evolution\footnote{$E(z)$ is the dimensionless Hubble parameter, approximated here as $E(z) = \sqrt{\Omega_m (1+z)^3 + \Omega_{\Lambda}}$ where $\Omega_m$ and $\Omega_{\Lambda}$ refer to their $z=0$ values. }:

\begin{equation}  L_{X\text{, bolometric}} = E(z)^{7/3}  \times L_{0\text{, bolo}} \left(\frac{M_{500}}{M_0}\right)^{\alpha} \end{equation}

However, we do not measure the bolometric luminosity. We have rest-frame 0.5-2.0 keV luminosities, so we can divide $L_{0\text{, bolo}}$ by a temperature-dependent bolometric correction $C_{\text{bolo}}$ and express this relation in terms of the 0.5-2.0 keV luminosity:

\begin{equation}  L_{X\text{, 0.5-2.0 keV}} = E(z)^{7/3} \times L_{0\text{, bolo}} \times C_{\text{bolo}}^{-1} \left(\frac{M_{500}}{M_0}\right)^{\alpha} \end{equation}

We set $M_0 = 4\times10^{14} M_{\odot}$. This equation then has two free parameters: $L_{0\text{, bolo}}$ (the normalization) and $\alpha$ (the slope). We fit for these parameters using forward-modeling. We create a grid of combinations of these two parameters. For each value of $\alpha$, we compute the effective halo mass using the procedure described in Appendix A. With the effective halo mass corresponding to each stellar mass bin, we can invert the $L_X$-$M_{500}$ relation to get an updated value of $L_X$. We then compare the model prediction of $L_X$ for each combination of parameters to the observed luminosities. The observed values of $L_X$ also depend slightly on the effective halo mass (through the $R_{500}$ aperture size), so we recompute the observed luminosities for each stellar mass bin using this new value of $R_{500}$ as well (this yields very small changes in the luminosity, of order a few hundredths of a dex). To minimize the effect of variations caused by the finite number of galaxies in the true catalog (i.e. the dispersion between different simulations of the same parameters, visible in Figure D.1) we perform 10 simulations of each bin for each combination of parameters and average the results. 

We compare each combination of parameters to the observed luminosities using the $\chi^2$ goodness of fit parameter. The uncertainty we use is the sum, in quadrature, of: the sample error on the mean for the data (i.e. the thin error bars in Figure 5, which are always larger than the thick error bars), the standard deviation of the best-fit luminosities in the ten simulated stacks, and an assumed 10\% uncertainty in the measured luminosity to account for uncertainties in the counts-to-flux conversion and the k-correction. We also include in quadrature a factor of 10\% in the bolometric correction applied to each bin; for the higher-mass objects (with larger bolometric corrections) this factor dominates the uncertainty budget. Our assumed values for the bolometric correction are listed in Table 2.

The resulting best-fit values are shown in the contour plot in Figure 6, along with $1\sigma$, $2\sigma$, and $3\sigma$ uncertainty regions (defined as $\chi^2 <$ 11.54,18.61,26.90, respectively). The best-fit combination of parameters is $\alpha = 1.85^{+0.15}_{-0.16}$, $L_{0\text{, bolo}} = 1.4\pm0.4\times10^{44}$ erg s$^{-1}$. The best-fit individual values, marginalized over the other parameter, are $\alpha = 1.84$, $L_{0\text{, bolo}} = 1.4\times10^{44}$ erg s$^{-1}$.

In Appendix F, we cross-check these results by fitting equation 5 to our binned data directly (using the effective value of $M_{500}$ for the bin) instead of forward-modeling the $L_X$-$M_{500}$ relation through our simulated catalog. We find agreement at the $2\sigma$ level, although the forward-modeled relation prefers a slightly lower slope. We also checked the results using the measurement errors (i.e. the thick error bars in Figure 5) instead of the sample error on the mean, and find essentially the same best-fit values for $\alpha$ and $L_{0\text{, bolo}}$, though with less acceptable $\chi^2$ values.

\begin{figure}
\begin{center}
\includegraphics[width=8.5cm]{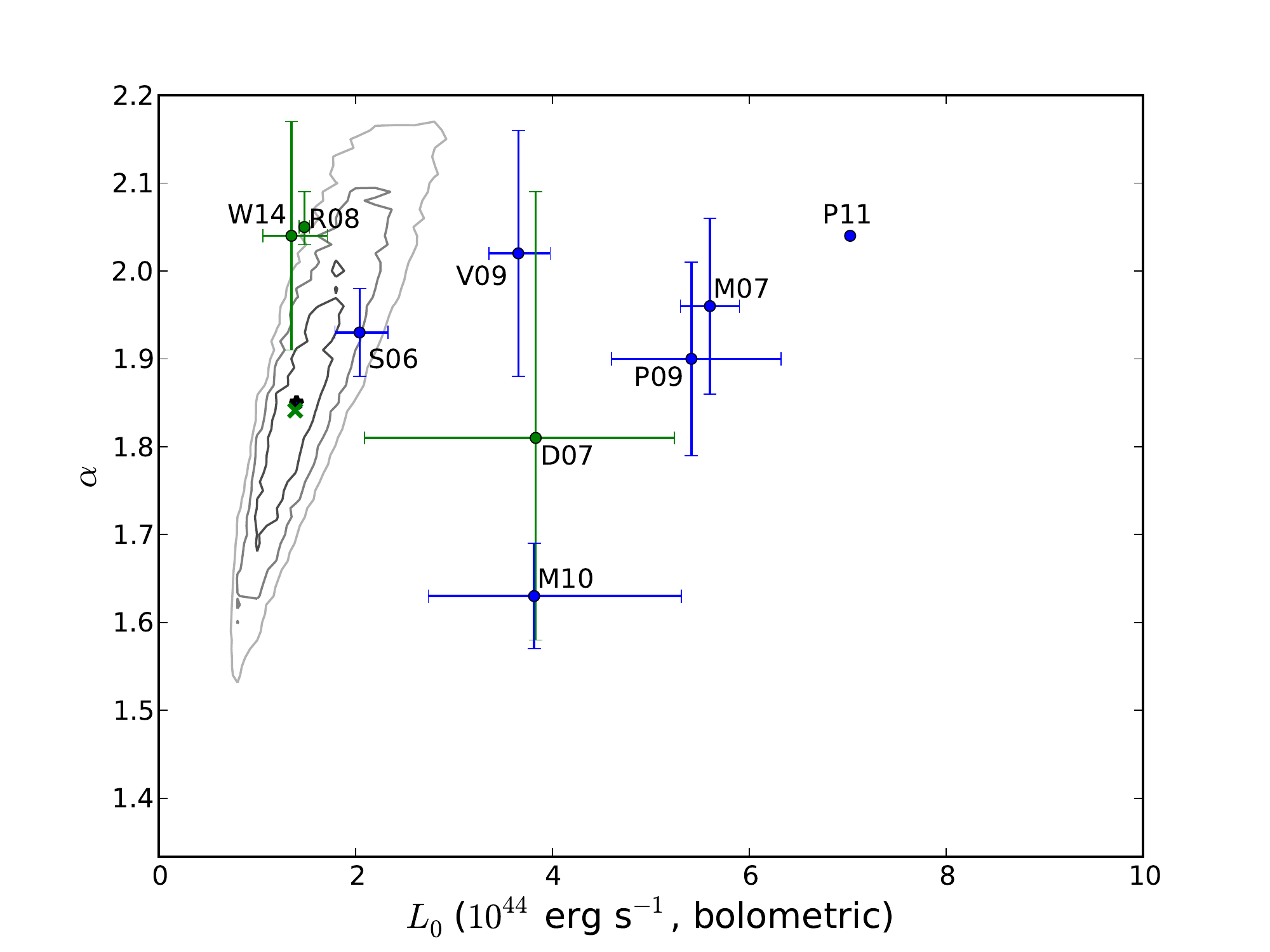}
\end{center}
\caption{Best-fit parameters for the $L_X$-$M_{500}$ relation, assuming the functional form of eq. (2) and normalizing to $M_0 = 4\times10^{14} M_{\odot}$. Contours indicate $1\sigma$, $2\sigma$, and $3\sigma$ confidence intervals. The best-fit combination of parameters is $\alpha = 1.85$, $L_{0\text{, bolo}} = 1.4\times10^{44}$ erg s$^{-1}$ (indicated with the asterisk). The best-fit individual values, marginalized over the other parameter, are $\alpha = 1.84$, $L_{0\text{, bolo}} = 1.4\times10^{44}$ erg s$^{-1}$ (indicated with the green 'X' symbol). For comparison, the blue and green points are adapted from other published works, as described in section 5 (blue corresponds to X-ray-selected samples, green to optical or near-IR selection). As described in the text, published results based on soft-band luminosities instead of bolometric luminosities will generally yield shallower slopes; in these cases we have adjusted the slope upwards to the value it would be expected to have if the relation were fit to the bolometric data.}
\end{figure}

In these relations we assume self-similar evolution since we consider it the simplest assumption. Moreover, \citet{Maughan2007} and \citet{Vikhlinin2009} both find that their samples are consistent with self-similar evolution, although this is generally a very difficult measurement to make and the exact nature of the evolution remains highly uncertain. Since $E(z)$ in our sample monotonically increases with luminosity (due to the optical flux limit in SDSS), changing the assumed parameterization of the redshift evolution will affect the inferred slope of the $L_X$-$M_{500}$ relation, but it will have little effect on the normalization.

\subsection{Other Measurements of the $L_X$-$M_{500}$ Relation}

There have been a number of previous estimates of the $L_X$-$M_{500}$ relation, all of which have been restricted to galaxy clusters and massive groups. In Figure 6 we have plotted a number of the more recent results, for comparison with our own results. These studies differ in sample selection, redshift range, X-ray energy band, and assumptions about the redshift evolution of the clusters. This makes a perfect comparison very difficult, but we attempt a simple version here to illustrate a few broad points. Each of the relations we consider fits their observations to a power-law similar to eqs. 4 and 5, although the relations often have different choices for the energy range, the exponent on the $E(z)$ factor, and the reference mass $M_0$. We evaluate each relation at $M_{500} = 4\times10^{14} M_{\odot}$ (which is the pivot point we use in our relation, and lies just above our highest-mass bin). We apply bolometric corrections (listed below) to the normalization of each relation, assuming a gas temperature of 5 keV (the approximate temperature of gas in hydrostatic equilibrium with a halo of our reference mass). The bolometric correction also has an effect on the slope: for a relation which is fit to the soft-band luminosity, an increasing fraction of the total X-ray emission falls outside the soft band and so the slope is underestimated. The exact correction to the slope depends on the energy band and redshift range; we estimate this correction for each relation based on an APEC model with $Z=0.4 Z_{\odot}$ and mean absorbing column of $5\times10^{20}$ cm$^{-2}$, using the assumption $M_{500} \propto T^{2/3}$ to convert between temperature and mass. The typical value of the increase to the slope is around 0.4. 

Finally, the evolutionary correction also has an effect on both the slope and normalization of the relation, although both of these effects are minor for the low redshifts examined in most of these studies. In Figure 6 we just set the $E(z)$ terms to unity, effectively evaluating each relation at $z=0$. This will introduce errors into our estimation of the slopes and normalizations, but the errors will be small. For a relation with a typical $z=0.1$, $E(z) \approx 1.05$ and so the different assumptions of self-similar evolution and no evolution lead to 10\% differences in the inferred normalization of the $L_X$-$M_{500}$ relation. 

The S06 data point corresponds to \citet{Stanek2006}, who give relations for $\Omega_m = 0.24$ and $\Omega_m=0.30$ cosmology. We compute the slope and normalization of each relation at our pivot point and take their mean to get the result for the WMAP7 cosmology. The S06 relations use $M_{200}$ as the independent variable instead of $M_{500}$; we adopt a value of $5.7\times10^{14} M_{\odot}$ for $M_{200}$, which assumes an NFW profile \citep{Navarro1997} with a concentration of 5.1 \citep{Prada2012}. We apply a bolometric correction of 2.3 to convert the normalization from their 0.1-2.4 keV band into bolometric luminosity. As discussed above, the slope of the relation is also bandpass-dependent; for this bandpass we correct the slope upwards by 0.40.

The M07 data point is \citet{Maughan2007}, which provides a bolometric spectroscopic luminosity measured within $R_{500}$, so no conversions are necessary.  The P09 data point is \citet{Pratt2009}, from which we use their Malmquist bias-corrected relation evaluated using BCES regression (again the bolometric relation is provided, so no correction is necessary). The V09 data point is \citet{Vikhlinin2009}, and we use the relation given in equation 22 from that paper. This relation has a pivot point at $1 M_{\odot}$, so extrapolating the uncertainties in the slope up to $4\times10^{14} M_{\odot}$ yields apparent uncertainties in the normalization which are orders of magnitude in size; we therefore neglect this uncertainty when computing the uncertainty in the normalization at our pivot point. For this relation we apply a bolometric correction of 2.9 to convert from 0.5-2.0 keV into bolometric luminosity. We also adjust the slope upwards by 0.41, which is the approximate conversion for a powerlaw measured the 0.5-2.0 keV band to the bolometric form. The M10 data point is \citet{Mantz2010}, from which we use the bolometric $L$-$M$ relation for the full dataset. The P11 data point is \citet{Planck2011}, from which we use their fiducial relation, which has no correction for Malmquist bias. For this relation we also apply a bolometric correction factor of 2.3 to the normalization and adjust the slope upwards by 0.40.

Finally, we also examine three samples of clusters identified from optical or near-IR photometry, unlike all the X-ray flux-limited relations listed above. The D07 data point is \citet{Dai2007}, which uses a stacking approach very similar to ours, applied to galaxy clusters identified in the 2MASS catalog. We use the bolometric luminosities and mean redshifts for their five mass bins, and the estimates of $M_{500}$ from \citealt{Dai2010}, and fit relations of the form of equation 4 to these five data points in order to estimate the slope and normalization at our pivot point. R08 is \citet{Rykoff2008}, which uses 17000 galaxy clusters from the maxBCG catalog \citep{Koester2007} selected from SDSS. They also measure the X-ray luminosities from ROSAT stacking, and use optical richness (calibrating with weak lensing) as a mass proxy. We shift their relation to our pivot point, and we also convert $M_{200}$ into $M_{500}$ and apply a bolometric correction of 2.0 to the normalization and a correction of 0.39 to the slope (corresponding the 0.5-2.0 keV band at $z=0.25$). Finally, W14 is \citet{Wang2014}, which also uses an optically-selected sample of clusters from SDSS. We use their $L_X$-$M_{200}$ relation (their equation 11), and again convert $M_{200}$ into $M_{500}$, apply a bolometric correction of 2.3 to the normalization, and adjust the slope upwards by 0.40.

The most striking feature of these relations is the huge variation among them. While the slopes are generally clustered around 1.8-2.1 (after the correction for the soft-band relations), the normalizations span a very large range of values. As discussed above, some of this disagreement is due to differences in assumptions about $E(z)$ and the different redshift ranges examined by each study, but given the low redshifts of most of these objects this is very unlikely to explain the bulk of the discrepancy. Rozo et al. (2014) have studied the discrepancies in X-ray scaling relations in a systematic way, showing that there are persistent disagreements between various $L_X$-$M$ relations which do not depend on the evolution factor. One significant issue is the assumed $M$-$f_{\text{gas}}$ relation, which is responsible for the difference in normalization between M10 and P09/V09. Other systematic offsets in the X-ray observables are also implicated, and several correction factors are required in order to bring these three relations into agreement. Even after applying these offsets, \citet{Rozo2014} note that extending these relations to lower masses (which is the area of interest for our analysis) will lead to additional divergences.

\subsection{Normalization and Scatter in the $L_X$-$M_{500}$ relation}

The other striking feature of Figure 6 is the difference in normalization between the optically-selected samples (ours, D07, and W14) and the X-ray selected samples (S06, M07, P09, V09, M10). Except for S06, the X-ray-selected samples all have higher normalizations than the optically-selected samples. The obvious culprit is Malmquist bias: there is intrinsic scatter in $L_X$ at a given $M_{500}$, and any sample of clusters which includes X-ray flux in the selection criterion (as all of the above X-ray samples do) will systematically overestimate the normalization.

However, each of the above X-ray studies has already attempted to account for Malmquist bias. \citet{Stanek2006} infer an intrinsic scatter in $L_X$ at a fixed halo mass of $\sigma_{\text{ln L}} = 0.68$. This is much larger than the intrinsic scatter inferred by \citet{Maughan2007} ($\sigma_{\text{ln L}} = 0.17-0.39$), \citet{Pratt2009} ($\sigma_{\text{ln L}} = 0.38$), \citet{Vikhlinin2009} ($\sigma_{\text{ln L}} = 0.39$), or \citet{Mantz2010} ($\sigma_{\text{ln L}} = 0.19$). This is a wide range in the inferred magnitude of the scatter. In theory, a logarithmic scatter $\sigma_{\text{ln L}} \approx 0.4$ should produce a natural logarithmic bias of order $\alpha \times 0.4^2 \approx 0.3$, corresponding to an offset of about 30\%. This could explain some of the observed offset between the optically-selected samples and the X-ray selected samples, but probably not all of it. On the other hand, if the Stanek value of 0.68 is correct, then the bias rises to about 80\%, which brings most of the relations within $1\sigma$ of our results. Depending on the scatter, it is conceivable that the correction for Malmquist bias is the dominant cause of the differences in normalizations in Figure 6, although it is by no means the only possible cause.

For comparison, we can convert our measurements of the standard error on the mean, derived from bootstrapping analysis, into estimates of the standard deviation in $L_X$ for individual galaxies by multiplying the standard error by the square root of the number of galaxies in each bin. These values are shown in Appendix G. We find a very significant amount of intrinsic scatter among our sample (more than enough to explain the offset in normalization), and also find that the scatter increases as the mass decreases towards galaxy groups and galaxies. 

There are several other reasons to suspect the intrinsic bias may be underestimated in many of the X-ray-selected studies. Observations of an optically-selected sample of galaxy groups \citep{Rasmussen2006} raised this concern several years ago. Simulations by \citet{Rasia2012} predict underestimates in the mass of order 25-35\% from X-ray measurements, and \citet{Sereno2014} report biases of 40\% in mass between hydrostatic X-ray estimates and weak lensing. Moreover, a study by \citet{Hicks2013} finds that optically selected clusters tend to be have more disturbed X-ray morphologies than the clusters in typical X-ray samples, which suggests that typical X-ray samples may be underestimating the scatter in representative samples of clusters. \citet{Angulo2012} recently attempted to estimate the expected intrinsic scatter in the  $L_X$-$M_{500}$ relation, predicting a scatter in log$_{10}$ M at fixed $L_X$ of 0.28. Given their predicted slope of 1.5, this translates to $\sigma_{\text{ln L}} \approx 0.98$ (which is comparable to our measurement). Finally, P13 find a $Y$-$M$ relation with a normalization 20\% lower than is found in X-ray-selected samples, even for the most massive clusters.

As an independent way of checking the importance of selection effects, in Appendix H we impose an X-ray flux limit of $1\times10^{-12}$ erg s$^{-1}$ cm$^{-2}$ on our sample. This yields a much noisier and poorer subsample, but the inferred normalization of the $L_X$-$M_{500}$ relation rises by more than a factor of two, making it consistent with other X-ray flux-limited studies.

\subsubsection{Malmquist Bias in Locally Brightest Galaxies?}

The preceding discussion is premised upon the claim that Malmquist bias in the $L_X$-$M_{500}$ relation should be less significant for optically-selected samples than for X-ray-selected samples. This is plausible, since variation in optical luminosity of galaxies is not expected to be strongly linked to variations in the properties of the hot halo gas. However, we can also show this explicitly by generating a volume-weighted stack of our locally brightest galaxies. To do this, we take the k-corrected $r$-band magnitude for each locally brightest galaxy, and weight each RASS image and each exposure map by the maximum volume within which this galaxy would be detected. The results are nearly identical to the values in Table 3 for our fiducial run; the median offset between the Malmquist-corrected luminosities and the fiducial (unewighted) luminosities is just 0.02 dex and the largest offset is 0.14 dex (in the bin with log $M_{\ast}/M_{\odot}$ = 10.1-10.2; in the upper twelve bins the largest offset is just 0.08 dex). These offsets are comparable to the $1\sigma$ uncertainties on $L_X$ (see Table 3). This justifies our claim that the $L_X$-$M_{500}$ relation of our optically-selected sample is much less significantly affected by Malmquist bias than the relation for the X-ray-selected samples.

\subsection{Implications for AGN Feedback}

As noted in section 5.1, the observed slope of the $L_X$-$M_{500}$ relation is always much steeper than the self-similar prediction of 4/3.  If we assume the steepness of the observed slope is due to nongravitational heating, then AGN feedback is likely an important contributor. In this section we assume that AGN feedback controls the behavior of the hot gas in these systems, and we discuss what conclusions can be drawn from the data.

A useful framework for this discussion is the distinction between two classes of AGN feedback studied in \citet{Gaspari2014} using 3D high-resolution hydrodynamic simulations (and summarized briefly in Section 1).  In ``self-regulated'' models, thermal instabilities due to radiative cooling in the hot halo lead to steady condensation onto the central black hole, which boosts the accretion rate up to 100 times over the Bondi rate, re-heating the central region through the injection of mechanical energy and restarting the cycle. In ``thermal blast'' models this cycle is much more extreme, with steady cold gas inflow triggering quasar-mode feedback near the Eddington limit.

\citet{Gaspari2014} showed that models of thermal blast type (i.e. violent feedback) generically produce a break in the $L_X$-$T$ relation at $T \sim 1$ keV, below which point the thermal heating overcomes the binding energy of the hot gas. This type of feedback destroys the cool core and overheats the hot gas halo, increasing the cooling time to significantly above the Hubble time. On the other hand, more modest self-regulated AGN feedback does not generate such a break since the heating and cooling remain roughly balanced at all mass scales for several Gyr. In order to compare our data to these simulations, we have converted the effective $M_{500}$ values into effective temperatures, using the Sun et al. (2009) $M_{500}$-$T$ relation (these values are also listed in Table 2). 

Figure 7 shows the results of this comparison, both for $L_{\text{total}}$ and for $L_{\text{CGM}}$. Lacking measurements, we assume the same temperature for both regions, which is an oversimplification although in most cases the difference is only a few percent (see \citealt{Gaspari2014}). The difference between the X-ray flux-limited sample and the full sample is discussed in Appendix H. Since this plot is essentially a rescaled version of the $L_X$-$M_{500}$ plot the same offset in normalization is visible for our optically-selected sample. 

The slope of our $L_X$-$T$ relation is steeper than the self-similar prediction (a slope of 2 when thermal bremsstrahlung dominates the emission) for the total sample. This shows the same effects of non-gravitational heating as the $L_X$-$M_{500}$ relation. There is also clearly no break in the $L_X$-$T$ relation, which is evidence for AGN feedback operating through a gentle self-regulated mechanism instead of through more powerful thermal blasts. As suggested by Mulchaey (2000) and others, the break which is sometimes seen in other samples may be due to the low sensitivity of current instruments for observing single groups, which our stacking largely circumvents.

\begin{figure}
\begin{center}
\subfigure[$L_{\text{total}}$]{\includegraphics[width=8cm]{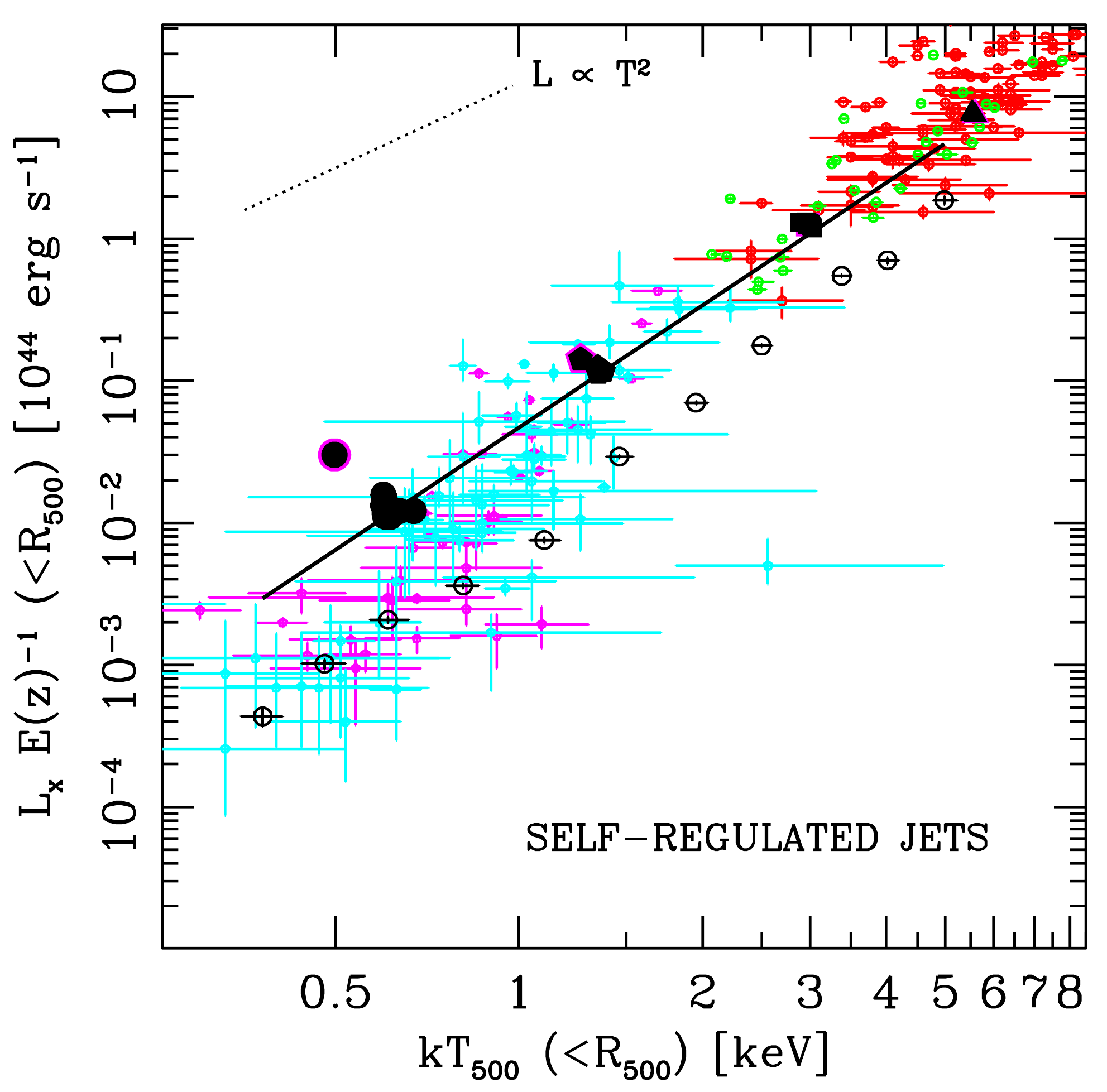}}
\subfigure[$L_{\text{CGM}}$]{\includegraphics[width=8cm]{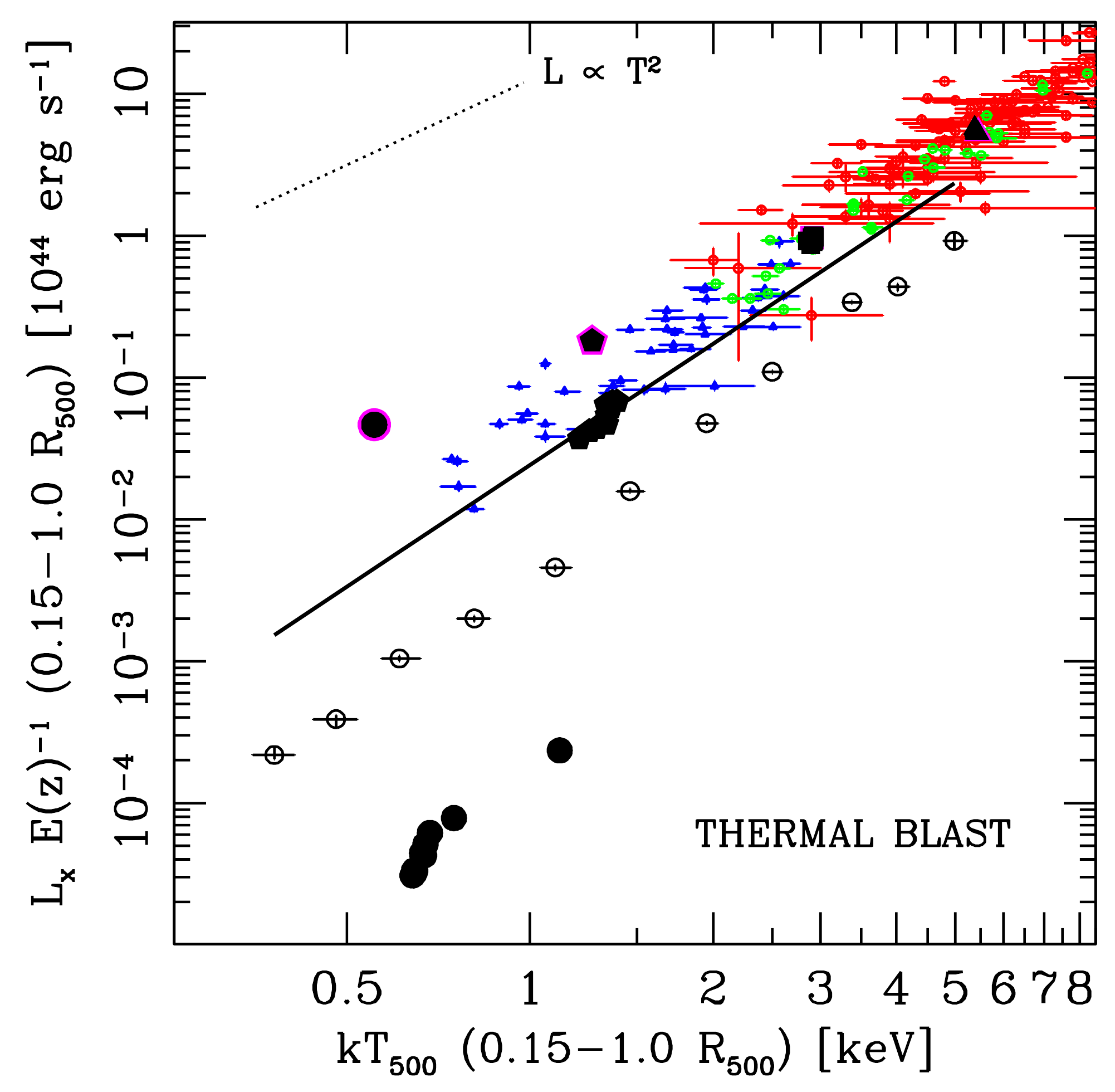}}
\end{center}
\caption{Luminosity-temperature relation for galaxies, groups, and clusters. The upper plot shows $L_{\text{total}}$ measured for our full sample of LBGs (open data points) as well as the best-fit relation to the X-ray flux-limited sample of LBGs (black line). The lower plot shows our $L_{\text{CGM}}$ results plotted in the same way, using the same assumed temperatures for the CGM annulus as for the total system. In both plots, the data are compared to the indicated AGN feedback models as simulated by \citet{Gaspari2014}; for each system 5 Gyr of evolution are plotted, separated into 500 Myr time steps (starting with the point outlined in magenta). The self-regulated feedback models (upper panel) have been rescaled by realistic gas fractions described in the text. For comparison, measurements of individual objects are also included, from \citet[][cyan]{Helsdon2000b, Helsdon2000a}, \citet{Mulchaey2003} and \citet[][magenta]{Osmond2004}, \citet[][blue]{Sun2009}, \citet[][green]{Pratt2009}, and \citet[][red]{Maughan2012}. The predictions for self-regulated feedback match the flux-limited observations well, while thermal blasts predict a break around 1 keV which is not observed in our data. }
\end{figure}

We can make this comparison more quantitative as well. In Figure 7a, the observed values of $L_{\text{total}}$ are compared to the predictions for self-regulated feedback from \citet{Gaspari2014}. The simulations use initial conditions corresponding to gas-rich X-ray bright systems with active AGN feedback, so the X-ray flux-limited sample is the most relevant sample for comparison. Four simulations are considered, with $M_{200}$ at $z=0$ of approximately $2\times10^{13} M_{\odot}$, $7\times10^{13} M_{\odot}$, $3\times10^{14} M_{\odot}$, and $8\times10^{14} M_{\odot}$, corresponding to poor and massive groups and clusters, respectively. The original setup used $f_{\text{gas}} \simeq 0.15$ for all systems to test if self-regulated feedback could provide any strong evacuation ab initio, but realistic systems have lower gas fractions at lower mass, so we rescaled the three lower-mass self-regulated simulations (upper panel) adopting realistic gas fractions of 0.14, 0.10, and 0.07 respectively\footnote{Applying the same procedure to the ``thermal blast'' simulations would make the discrepancy with observations worse, so we neglect that rescaling for these simulations.}.

The consistency between observations and gently self-regulated simulations is good across the entire mass range. The key result is that self-regulated mechanical AGN feedback can preserve the $L_X$-$T$ relation (excising the core or not) without any major break for several Gyr. In order to preserve the slope at all masses, the simulations require tight self-regulation based on the cooling of the central gas instead of inefficient hot Bondi-like accretion (\citealt{Gaspari2013}, \citealt{Gaspari2014b}). As in our observed sample, the scatter increases in the regime of less massive groups (Appendix G). 

Figure 7b compares the observed values of $L_{\text{CGM}}$ to the \citet{Gaspari2014} predictions for the powerful ``thermal blast'' feedback. This class of AGN feedback creates a strong break in the scaling relations below 1 keV. The system is nearly emptied and even the core-excised luminosity $L_{\text{CGM}}$ is seen to be two or three dex below the measured values. This discrepancy is a defining characteristic of strong thermal blast models: the central cooling time is raised well above the Hubble time when a thermal blast occurs, transforming the systems into non-cool-core objects. In flux-limited samples, this is not observed: the majority of detected systems harbor a weak or strong cool core (\citealt{Mittal2009}, \citealt{Sun2009}).

Overall, the power-law nature of our observed $L$-$T$ relations is very difficult to explain with thermal blasts, and it indicates that gentle self-regulated mechanical feedback can preserve the large-scale scaling relations from massive clusters to L* galaxies. More detailed comparisons between observations and new simulations (e.g., varying $f_{\text{gas}}$) will be carried out in future work.

\section{Galaxy-Scale Halos}

In the regime of massive (elliptical) galaxies, it is common to examine scaling relations between $L_X$ and a stellar mass proxy such as $L_K$ or $L_B$, rather than relating $L_X$ to the halo mass. This is a much more straightforward relation to fit to our data than the $L_X$-$M_{500}$ relation, since our independent variable is already $M_{\ast}$. Here we explore linear fits to the data in Fig.4 for relations with the form

\begin{equation}  L_{X\text{, 0.5-2.0 keV}} =  C_{\text{bolo}}^{-1} \times L_{0\text{,* bolo}} \left(\frac{M_{*}}{M_{*,0}}\right)^{\alpha} \end{equation}

\noindent as well as relations without the bolometric correction, i.e.

\begin{equation}  L_{X\text{, 0.5-2.0 keV}} = L_{0\text{,*}} \left(\frac{M_{*}}{M_{*,0}}\right)^{\alpha} \end{equation}

Since we are able to recover input X-ray luminosities well, there is no need to forward-model these relations through the simulated galaxies in this case. The forward modeling allows us to constrain uncertainties in the matching between $M_{\ast}$ and $M_{500}$, but these relations use $M_{\ast}$ as the independent variable, so we can just fit to the observed data directly.  The error budget is identical to the budget in Section 4. For the relations with a bolometric correction, we include a 10\% uncertainty in this factor as well, and this increases the magnitude of the uncertainty and reduces the $\chi^2$. 

We also explore the effect of adding the stellar mass of the satellite galaxies into the relation as well. To estimate this, we use the conditional mass function inferred from the abundance-matching simulations of \citet{Moster2010}. The satellite galaxies contain 13\% as much stellar mass as the central galaxy in the lowest-mass bin (log $M_{\ast}$ = 10.8-10.9), rising smoothly up to 232\% as much stellar mass as the central galaxy in the highest-mass bin. 

In Table 4 we present each of these best-fit relations. The normalization of these relations is barely affected by the inclusion of satellite galaxies, since our pivot point is at a total stellar mass of $10^{11} M_{\odot}$ where satellite galaxies only contain 24\% as much stellar mass on average as the cental galaxy. Accounting for satellite galaxies does lower the slope significantly, though. Excluding the core region lowers the normalization but leaves the slope of the relation largely unaffected. Finally, converting to the bolometric band steepens the inferred slope. The best-fit regions for the bolometric form of these relations (including stellar mass in subhalos) are shown in Figure 8. 

Note that these formulations assume all the emission we observe is from hot gas. We showed in section 3 that this is likely to be true for log $M_{\ast} >$ 10.8, but we can also test this by excluding the lowest two data points and re-fitting the other 10 data points. We find that the best-fit values are unchanged within the $1\sigma$ uncertainties, suggesting that our results are not strongly affected by contamination in these two bins.

\begin{figure}
\begin{center}
\includegraphics[width=8.5cm]{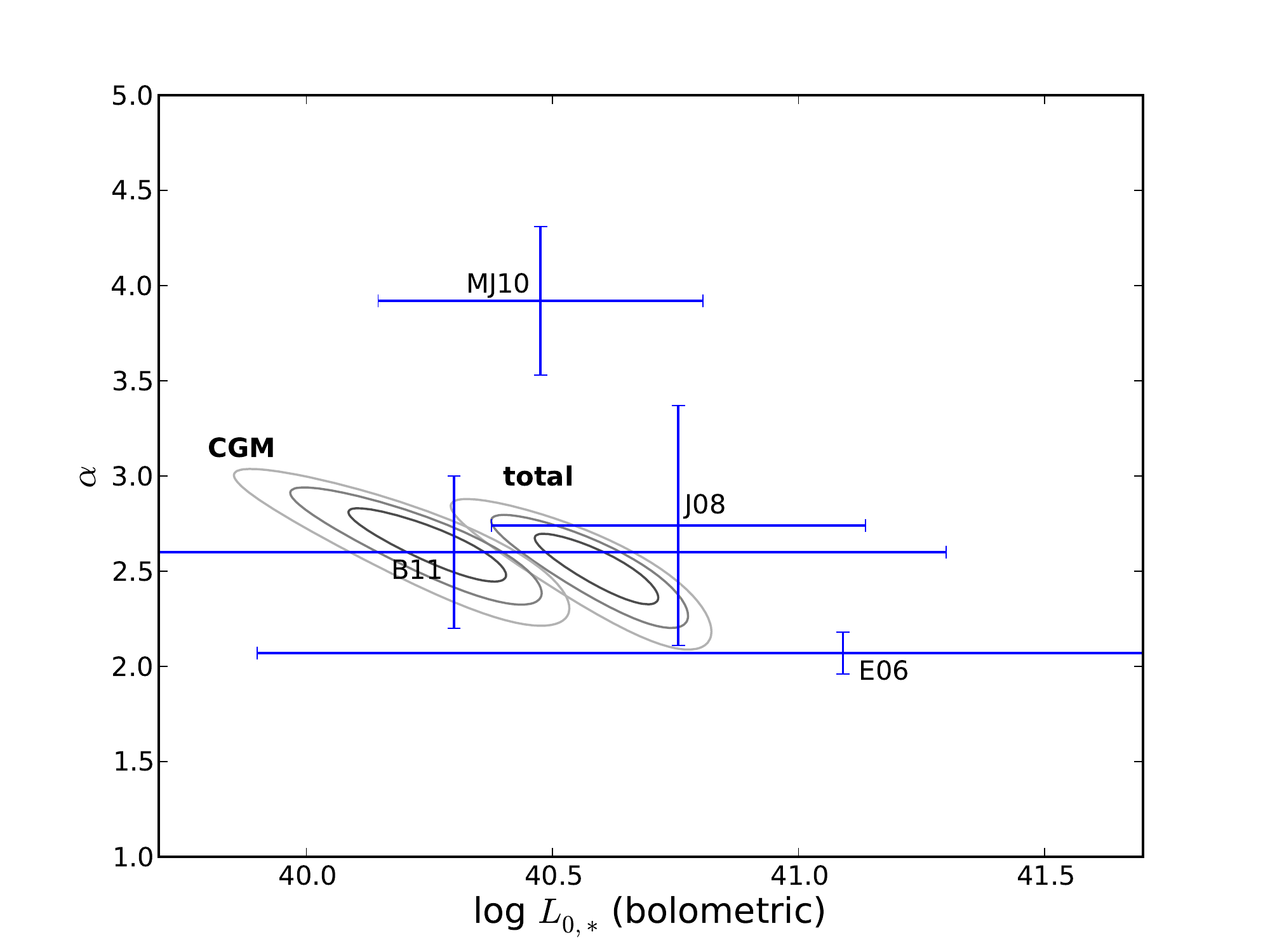}
\end{center}
\caption{Best-fit values for the free parameters in the $M_{\ast}$-$L_X$ relation (equation 6), including both the stellar mass of the central galaxy and the satellite galaxies within $M_{\ast}$. Results are fit both to $L_{\text{total}}$ (which has the higher normalization) and $L_{\text{CGM}}$ (which has the lower normalization). Contours indicate $1\sigma$, $2\sigma$, and $3\sigma$ uncertainty regions for each relation and error bars. Approximate comparisons are also made with results from the literature, as described in the text. }
\end{figure}

\begin{table*}
\begin{minipage}{92mm}
\caption{$L_X$ - $M_{\ast}$ relations }
\begin{tabular}{lccc}
\hline
Relation & log $L_0$ & $\alpha$ & $\chi^2$ / d.o.f.\\
& (erg s$^{-1}$) & &\\
\hline
$L_{\text{total}}$, 0.5-2.0 keV, central only & $40.75$ & $3.34$ & 1.19\\
$L_{\text{total}}$, 0.5-2.0 keV, with satellites & $40.63$ & $2.21$ & 1.16\\
$L_{\text{CGM}}$, 0.5-2.0 keV, central only & $40.45$ & $3.46$ & 1.44\\
$L_{\text{CGM}}$, 0.5-2.0 keV, with satellites & $40.32$ & $2.29$ & 1.72\\
$L_{\text{total}}$, Bolometric, central only & $40.71$ & $3.81$ & 1.27\\
$L_{\text{total}}$, Bolometric, with satellites & $40.60$ & $2.53$ & 0.68\\
$L_{\text{CGM}}$, Bolometric, central only & $40.38$ & $4.02$ & 0.74\\
$L_{\text{CGM}}$, Bolometric, with satellites & $40.26$ & $2.64$ & 0.66\\
\hline
\end{tabular}
\\
\small{ Best-fit parameters for the $L_{\text{total}}$ - $M_{\ast}$ relation for massive galaxies (eqs. 4 and 5). We examine eight different relations, described in section 6. The reduced $\chi^2$ of the best fit is also indicated (each fit has 10 degrees of freedom). In general the fit is improved by the inclusion of stellar mass from satellite galaxies, and the bolometric relations also have better fits due to extra uncertainty from the bolometric corrections. }
\end{minipage}
\end{table*}

These results can be compared to existing $L_X$-$L_K$ relations for individual elliptical galaxies (e.g. \citealt{Ellis2006}, \citealt{Jeltema2008}, \citealt{Mulchaey2010}, \citealt{Boroson2011}). In order to make this comparison, it is necessary to assume a parametric form for the mass-to-light ratio as a function of either $L_K$ or $M_{\ast}$. We use the \citet{DeLucia2007} semi-analytic prescription to estimate the K-band M/L ratio for central galaxies at $z=0.1$ with stellar masses around $10^{11} M_{\odot}$, and infer a M/L ratio of 0.41. This prescription assumes a \citet{Chabrier2003} initial mass function; using a \citet{Salpeter1955} initial mass function would increase the inferred stellar mass by nearly a factor of two. Other uncertainties in the stellar mass exist as well, such as potential underestimates of the stellar light at large radii \citep{Kravtsov2014}, although the effect of this sort of error is bracketed by the ``central only'' and ``with satellites'' cases examined above, suggesting that it does not qualitatively change our results.

Using our simple estimate of the K-band M/L ratio, we then evaluate each of the above relations at a luminosity of $L_K = 2.4\times10^{11} L_{\odot}$ (corresponding to $M_{\ast} = 10^{11} M_{\odot}$) to estimate their normalization at our pivot point. The error bars on the parameters of these relations in Figure 8 are approximate $1\sigma$ uncertainties.

For \citet{Ellis2006}, we use their $K_S$-band relation fit with the expectation maximization algorithm to their largest combined dataset. For \citet{Jeltema2008}, we use their relation for group galaxies, excluding nondetections. We use the equation listed in section 3 for the \citet{Mulchaey2010} relation, which is fit to field galaxies. Both \citet{Jeltema2008}  and \citet{Mulchaey2010} are measured in the 0.5-2.0 keV band, so we apply our bolometric correction to convert to bolometric luminosities. Finally, for \citet{Boroson2011} we use their quoted slope of $2.6\pm0.4$, but they do not quote a normalization so we estimate the normalization from their Figure 5a (the uncertainties on this value span most of the plot, so the exact value is not important). We also apply a small bolometric correction of 1.2 from their 0.3-8.0 band. 

There are still some discrepancies between different $L_X$-$M_{\ast}$ relations, but these discrepancies can largely be attributed to different sample selection criteria (e.g. environment). Our relations seem completely consistent with most previous observations.

\section{Unifying X-ray and SZ Analyses}

P13 showed that the stacked effective Compton Y-parameter $Y_{500}$, as measured through the SZ effect, also has an unbroken power-law dependence on $M_{\ast}$ for locally brightest galaxies. Matching each value of $M_{\ast}$ to an effective value of $M_{500}$, P13 find a $Y_{500}$-$M_{500}$ relation with a slope close to the self-similar prediction of $5/3$. As discussed in Section 5, it is probably more appropriate to compute the slope of the $Y_{500}$-$M_{500}$ using forward-modeling. P13 note that fixing the slope of their relation at 5/3 yields an unacceptable reduced $\chi^2$ of 3, so it would be interesting to repeat their analysis using forward-modeling to understand the robustness of their result. Appendix F suggests that the results do not change dramatically if we use their simpler technique instead, although the slope changes a bit and the $\chi^2$ of the fit is poorer; however this may not necessarily remain true for an SZ analysis. In this section, we take the P13 result at face value, and offer brief comments on the implications. 

The P13 result implies self-similar scaling in the integrated gas pressure around central galaxies, while we find steeper scaling in the X-ray luminosity. Both analyses see no evidence of a break in the relation, down to at least log $M_{\ast} = 11.2$ for P13 and down to at least log $M_{\ast} = 10.8$ in this work. If both of these results are correct, it implies that the density profile of the hot gas within the halo is systematically flattening as the stellar mass of the central galaxy decreases. Thus, galaxy-sized halos would contain the cosmic fraction of baryons, just like galaxy clusters, but they would be less concentrated and therefore emit less X-ray radiation than their larger counterparts. 

This result can naturally be achieved with AGN feedback and/or with pre-heating, but definitive conclusions would  require a joint X-ray and SZ analysis of these galaxies, which may be performed in future work.

\section{Conclusions}

In this paper we have presented a stacking analysis of 201011 locally brightest galaxies, which are overwhelmingly centrals in dark matter halos. The masses of these dark matter halos range from below the mass of the Milky Way up to medium-sized galaxy clusters. For 20 logarithmically spaced bins in central galaxy stellar mass, we have computed the effective $M_{500}$ and $R_{500}$, and stacked the X-ray emission from these locally brightest galaxies as seen in the ROSAT All-Sky Survey.

Our analysis is novel in a few ways. It systematically examines the X-ray properties of halos across nearly three orders of magnitude in halo mass, generating a uniformly-calibrated dataset of X-ray luminosity across this whole range. Unlike most other studies of these X-ray properties, it relies on an optically-selected sample of central galaxies, and is therefore not subject to the Malmquist bias which typically plagues these analyses. Finally, it relies on the exact same galaxies as examined by P13 in the Sunyaev-Zel'dovich effect, which facilitates a comparison with these results.

The major results of our analysis  can be summarized as follows:

1. Extremely bright soft X-ray sources (i.e. AGN) are very uncommon in locally brightest galaxies, comprising no more than about $0.2\%$ of the sample (see Appendix B for more details).

2. In halos with central galaxies with log $M_{\ast}$ = 10.8-12.0, the 0.5-2.0 keV observed X-ray luminosity ($L_{\text{total}}$) can be related to the stellar mass of the central galaxy with a simple power-law function, with a slope of 3.34. This power-law holds for galaxies, galaxy groups, and galaxy clusters.

3. A simple power-law relation is also found after applying bolometric corrections to the data, and/or by adding additional stellar mass to each bin to account for stars in satellite galaxies. The bolometric correction tends to increase the slope, while accounting for satellite galaxies tends to decrease it. A power-law relation is also found if we exclude the emission from the central 15\% of $R_{500}$. These relations are consistent with previous measurements of the $L_X$-$L_K$ relation in elliptical galaxies, although we note that uncertainty remains due to the conversion from $L_K$ to $M_{\ast}$.

4. We clearly detect X-ray emission in the $(0.15-1)\times R_{500}$ annulus in galaxies with log $M_{\ast} = 10.9-11.0$, which is likely to be the signature of a hot gaseous halo, and we tentatively detect a similar signature in this annulus for galaxies with log $M_{\ast} = 10.8-10.9$.

5. We find statistically acceptable power-law fits for the $L_X$-$M_{500}$ relation for halos with central galaxies with log $M_{\ast}$ = 10.8-12.0. The best-fit slope (marginalized over the normalization) of this relation is $1.84$, which is steeper than the self-similar prediction of 4/3. This slope is consistent with other measurements, which have been restricted to galaxy clusters and massive groups instead of the much larger mass range studied here.

6. The inferred normalization of the $L_X$-$M_{500}$ relation from our data is lower than the normalizations inferred from X-ray-selected studies. Our normalization is consistent with the normalization inferred from a recent optically-selected sample, however, and we hypothesize that the X-ray-selected samples are not adequately accounting for the effects of Malmquist bias. 

7. To test this hypothesis, we show that the observed scatter in $L_X$ at a given mass is probably much larger than inferred by X-ray-flux selected studies. We show other evidence which also points to larger scatter and to increasing scatter for lower-mass halos. Finally, we stack an X-ray flux-limited subsample of locally brightest galaxies, and recover a higher normalization consistent with X-ray flux-selected samples.

8. The slope of our inferred $L_X$-$M_{500}$ relation is steeper than self-similar, showing the influence of non-gravitational heating (likely AGN feedback). Comparing our results to hydrodynamical simulations of AGN feedback performed by \citet{Gaspari2014}, we find excellent agreement with the predictions of gentle ``self-regulated'' mechanical feedback models, and significant disagreement with models using more violent ``thermal blast'' feedback. AGN feedback does not seem to break either the $L_X$-$M_{500}$ relation or the L-T relation.

9. Our results are an important complement to the results of P13. While the $Y_{\text{SZ}}$-$M_{500}$ relation may be self-similar, the $L_X$-$M_{500}$ relation is steeper than self-similar. These results may be reconciled by smoothly changing the density profile of the hot gas as a function of halo mass, although future work is needed to understand this in more detail.

\section{Acknowledgements}
The authors would like to thank J.A.R. Martin and the Planck Collaboration for graciously sharing their catalog of locally brightest galaxies. We would also like to thank J. Bregman, T. Davis, J.C. Hill, G. H{\"u}tsi, R. Khatri, I. McCarthy, R. Sunyaev, F. van den Bosch, and A. Vikhlinin for helpful discussions and advice during the production of this manuscript. This research has made use of data, software and/or web tools obtained from NASA's High Energy Astrophysics Science Archive Research Center (HEASARC), a service of Goddard Space Flight Center and the Smithsonian Astrophysical Observatory. This research has made use of NASA's Astrophysics Data System. This research made use of Astropy, a community-developed core Python package for Astronomy \citep{Astropy2013}.


\appendix
\counterwithin{figure}{section}

\section{Estimating Effective Halo Masses}

The independent variable in most of our analysis is the stellar mass of the locally brightest galaxy. This parameter can be used for analysis in some cases (such as the $L_X$-$L_K$ relation, see section 6) but for many purposes it is useful to relate the stellar mass of the LBG to the halo mass. We do this by defining an ''effective'' halo mass for each stellar mass bin, in a similar fashion to P13. 

For the twelve stellar mass bins which are dominated by hot gas, we calculate the effective halo mass through an iterative method. We assume a power-law $L_X$-$M_{500}$ relation (equation 5) which has two free parameters. In section 5 we explore a grid of parameters in order to find the best-fit values; for the rest of this paper we use the best-fit values which are $\alpha = 1.84$, $L_{0\text{, bolo}} = 1.4\times10^{44}$ erg s$^{-1}$. In each stellar mass bin, we generate 10 simulated stacked images by randomly drawing galaxies in the same stellar mass bin from the simulated catalog (described in section 3) and populating each galaxy with an X-ray halo using the halo mass and position within the halo of the simulated galaxy and the assumed $L_X$-$M_{500}$ relation (see Appendix D for more details). Each simulated stack contains the same number of simulated galaxies as are stacked in the observed catalog (these numbers are listed in Table 1). Each simulated galaxy is assigned the redshift of one of the observed galaxies.  We stack these simulated galaxies and then measure (using aperture photometry) the average $L_X$ within $R_{500}$ for each bin (assuming the effective halo mass used by P13 as an initial guess for $R_{500}$). We then invert the assumed $L_X$-$M_{500}$ relation in order to derive an effective $M_{500}$ for the bin. We explored the effect of iterating this procedure, since the initial guess for $R_{500}$ (used to measure $L_X$) can differ from the value implied by the inferred $M_{500}$, but we found that this difference is too small to have a significant effect on our results (especially since $R_{500}$ only depends on the cube root of $M_{500}$). This process yields converged estimates of $M_{500}$ for each stellar mass bin. We take the effective $M_{500}$ to be the mean value of $M_{500}$ across the ten realizations of this process for each stellar mass bin. The uncertainty on $M_{500}$ is the the standard deviation of these ten realizations. 

The assumption of a single power-law $L_X$-$M_{500}$ relation in estimating the masses has little effect on our results. The normalization of the relation cancels out and has no effect, and the slope only has a slight effect through the relative weighting of the different halo masses within each bin (see Table A1). Moreover, the use of a single power-law is justified by the results in Figure 5, which strongly suggest a single power-law relation applies for the uppermost 12 bins. 

For the other eight (low-mass) stellar mass bins, the emission is dominated by X-ray binaries instead of hot gas. It is therefore not appropriate to use the $L_X$-$M_{500}$ relation to estimate the effective halo mass. Since the emission is concentrated in the galaxy itself, we have no significant detections of X-ray emission in the CGM annulus, so the exact value of $R_{500}$ is not especially important. We therefore just invert the abundance matching relation of \citet{Moster2010} in order to compute the effective halo mass for each of these stellar mass bins. We propagate the uncertainties in this relation in order to estimate the uncertainties in the effective halo mass.

\begin{table}
\begin{center}
\caption{Effective Halo Masses from Simulated Stacks }
\begin{tabular}{cccc}
\hline
log $M_{\ast}$ & $M_{\text{eff, bf}}$ & $M_{\text{eff, ss}}$ & $M_{\text{eff, P13}}$\\
\hline
11.9-12.0 & 14.56 & 14.50 & 14.54 \\
11.8-11.9 & 14.41 & 14.33 & 14.34 \\
11.7-11.8 & 14.29 & 14.22 & 14.20 \\
11.6-11.7 & 14.08 & 14.01 & 13.99 \\
11.5-11.6 & 13.90 & 13.83 & 13.84 \\
11.4-11.5 & 13.70 & 13.62 & 13.63 \\
11.3-11.4 & 13.51 & 13.41 & 13.41 \\
11.2-11.3 & 13.29 & 13.19 & 13.21 \\
11.1-11.2 & 13.09 & 12.98 & 12.97 \\
11.0-11.1 & 12.91 & 12.80 & 12.71 \\
10.9-11.0 & 12.75 & 12.64 & 12.62 \\
10.8-10.9 & 12.60 & 12.47 & 12.40 \\
\hline
\end{tabular}
\\
\end{center}
\small{ Effective halo masses computed using the simulated catalog of locally brightest galaxies. $M_{\text{eff, bf}}$ are the values used in this work, and are computed using our best-fit $L_X$-$M_{500}$ relation. $M_{\text{eff, ss}}$ are computed assuming a self-similar ($\alpha = 4/3$)  $L_X$-$M_{500}$ relation; these values are not very different from our adopted values, which shows the relative insensitivity of the effective halo mass to the slope of the $L_X$-$M_{500}$ relation. For comparison, we also list the effective halo masses computed by P13 ($M_{\text{eff, P13}}$) for pressure-weighted observations instead of emission-weighted observations, assuming a self-similar scaling and using the \citet{Arnaud2010} pressure profile.  }
\end{table}

As an aside, we note that recent studies (\citealt{Kravtsov2014} and \citealt{Shankar2014}) have questioned the normalization and slope of the high end of the stellar mass - halo mass relation because of a possible underestimation by SDSS of the luminosity (and hence stellar mass) of bright elliptical galaxies. This issue has no effect on our analysis of the $L_X$-$M_{500}$ relation because the same SDSS stellar masses are used for our LBG sample as were used when adjusting the stellar mass function of the $\Lambda$CDM simulations to fit the observations. As a result, (monotonic) systematic errors in the SDSS stellar masses will match each other in the LBG catalog and in the simulated catalog, and will cancel out of our analysis (which in effect determines the stellar mass - halo mass relation by abundance matching).

\section{Contamination from Bright Sources}

Stacking analysis is a measure of the mean properties of a sample, so like any mean estimator it is fairly sensitive to outliers. In this section we examine our criteria for masking bright sources (point-like and extended) from our images to prevent them from biasing our results. 

We use essentially the same technique as in ABD13. This technique employs two separate filters. First, we cross-match each of our fields with the ROSAT Bright Source Catalog and Faint Source Catalog \citep{Voges1999} and mask any region which contains a bright source above a minimum count rate. We repopulate the masked region with photons that match the flux observed elsewhere in the field. Second, as a backup measure we discard any observation if it contains a pixel with more than 10 photons in it. 

For the first filter, if we set the minimum count rate to zero, we end up masking out many of the locally brightest galaxies themselves, especially at the high end where these galaxies lie at the centers of moderately luminous galaxy clusters or groups. We instead use a fairly conservative threshold, which is listed for each bin in Table B1. The corresponding luminosity is also listed for each threshold, if the sources to be masked were at the mean distance of the LBGs in that bin. These luminosities are more than an order of magnitude larger than our measured luminosities, and it is difficult to imagine any physically plausible way to produce these luminosities from hot gas. We therefore expect that the sources masked by this filter are a combination of bright foreground objects, background quasars, and AGN associated with the locally brightest galaxies themselves. The latter category is scientifically interesting, so for each bin we count the number of LBGs with a source above our minimum count rate  with a centroid within $R_{500}$. This number is also listed in Table B1, as well as the fraction of the total sample to which this number corresponds. In total there are 387 LBGs with masked sources within $R_{500}$, which is $0.2\%$ of the total. 

Note that masking these sources could slightly reduce the average inferred luminosity of our stacks, by a factor roughly proportional to the fractions listed in table B1, but these fractions are very low so the bias is insignificant.

We also want to test the effect of these two methods on our result. We therefore generate an alternative stack where we multiply the minimum count rate by two, a stack where we remove the second filter, and a stack where we make both changes at once. Table B2 shows the average inferred luminosities for each case. 

\begin{table}
\begin{center}
\caption{ Parameters for Masking Sources Listed in the ROSAT Bright and Faint Source Catalogs }
\begin{tabular}{ccccc}
\hline
log $M_{\ast}$ & min cps & log $\overline{L_{\text{eff}}}$  & $N_{500}$ & frac\\
 &  (count s$^{-1}$) & (erg s$^{-1}$) & & \\
\hline
11.9-12.0 & $2.8\times10^{-1}$ & 45.0 & 1 &0.0278\\
11.8-11.9 & $1.5\times10^{-1}$ & 44.7 &3 &0.0263\\
11.7-11.8 &  $1.0\times10^{-1}$& 44.4 &7 &0.0154\\
11.6-11.7 &  $7.7\times10^{-2}$& 44.1 &13&0.00980\\
11.5-11.6 &  $4.5\times10^{-2}$& 43.8 &28&0.00944\\
11.4-11.5 &  $2.8\times10^{-2}$& 43.5 &48&0.00804\\
11.3-11.4 &  $2.2\times10^{-2}$& 43.2 &54&0.00562\\
11.2-11.3 &  $1.3\times10^{-2}$& 42.9 &68&0.00479\\
11.1-11.2 &  $9.5\times10^{-3}$& 42.6 &45&0.00244\\
11.0-11.1 &  $5.0\times10^{-3}$& 42.3 &27&0.00125\\
10.9-11.0 &  $3.2\times10^{-3}$& 42.0 &25&0.00110\\
10.8-10.9 &  $1.7\times10^{-3}$& 41.7 &23&0.00102\\
10.7-10.8 & $1.1\times10^{-3}$ & 41.5 &13&0.00065\\
10.6-10.7 & $1.3\times10^{-3}$ & 41.5 &18&0.00105\\
10.5-10.6 & $1.4\times10^{-3}$ & 41.5 &8&0.00060\\
10.4-10.5 & $1.6\times10^{-3}$ & 41.5 &2&0.00019\\
10.3-10.4 & $2.0\times10^{-3}$ & 41.5 &2&0.00027\\
10.2-10.3 & $2.2\times10^{-3}$ & 41.5 &2&0.00035\\
10.1-10.2 & $2.4\times10^{-3}$ & 41.5 &0&0\\
10.0-10.1 & $2.6\times10^{-3}$ & 41.5 &0&0\\
\hline
\end{tabular}
\\
\end{center}
\small{Numbers used for masking sources listed in the ROSAT Bright and Faint source catalogs. The min cps column lists the minimum count rates (as listed in the catalogs) for each bin; any source with a listed count rate higher than this value is masked if it falls anywhere within the stacked fields. The $\overline{L_{\text{eff}}}$ column shows the luminosity of a source at the mean distance of the LBGs in each bin, if it had the minimum count rate required to be masked. These luminosities are much larger than plausible hot gas luminosities for each bin. $N_{500}$ is the number of LBGs in each bin with masked source that have a centroid within $R_{500}$, and frac is the fraction of the stacked LBGs in each bin with such a source.  }
\end{table}

\begin{table}
\begin{center}
\caption{ Average $L_X$ for Different Point Source Masking Techniques }
\begin{tabular}{ccccc}
\hline
log $M_{\ast}$ & log $L_X$ & log $L_X$ & log $L_X$ & log $L_X$\\
& (fiducial) & (2$\times$ min cps) & (filter B off) & (both)\\
\hline
11.9-12.0 & 43.82 & 43.82 &43.68& 43.68\\
11.8-11.9 & 43.46 & 43.46 &43.46& 43.46\\
11.7-11.8 & 43.39 & 43.39 &43.39& 43.39\\
11.6-11.7 & 42.98 & 42.98 &43.98& 42.98\\
11.5-11.6 & 42.64 & 42.65 &42.64& 42.65\\
11.4-11.5 & 42.34 & 42.35 &42.34&42.35\\
11.3-11.4 & 41.80 & 41.80 &41.80&41.80\\
11.2-11.3 & 41.52 & 41.52 &41.52&41.52\\
11.1-11.2 & 41.29 & 41.29 &41.27&41.27\\
11.0-11.1 & 40.97 & 40.97 &40.92&40.91\\
10.9-11.0 & 40.58 & 40.58 &40.53&40.53\\
10.8-10.9 &  40.40 &40.39 &40.40&40.39\\
10.7-10.8 & 39.96 &39.96& 39.74&39.73\\
10.6-10.7 & 40.10 &40.11 &40.07&40.08\\
10.5-10.6 & 39.60 &39.59 &39.60&39.59\\
10.4-10.5 & 38.96 &39.03 &$<0$&$<0$\\
10.3-10.4 & 39.93 &39.93 &39.91&39.90\\
10.2-10.3 & 40.00 &40.02 &39.76&39.79\\
10.1-10.2 & 39.60 &39.60 &39.46&39.46\\
10.0-10.1 & $<0$ & $<0$ &$<0$&$<0$\\
\hline
\end{tabular}
\\
\end{center}
\small{Inferred $L_{\text{total}}$ for each bin, using variations on our fiducial point source masking technique. The fiducial technique masks sources from the ROSAT Bright and Faint Source Catalogs above a minimum count rate and excludes any observation containing a pixel with more than 10 counts (Filter B). The effects of modifying one or both of these filters are shown in the final three columns. For the upper 12 bins where we attribute the X-ray emission to hot gas, our results are quite robust to the details of the masking technique.  }
\end{table}

As Table B1 shows, above log $M_{\ast} = 10.7$-10.8, the differences in the average luminosity are insignificant between the different techniques, except for a small difference in the uppermost bin. In fact, if we turn off one or both filters, the luminosity is about as likely to decrease as to increase, which suggests that about half the bright sources we are masking lie outside $R_{500}$ and are contributing their photons to the background region instead of the source region. Below log $M_{\ast} = 10.8$-10.9, the differences in the average luminosity become more important, suggesting that a significant portion of the inferred signal is coming from individual objects. The most extreme example of this is the log $M_{\ast} = 10.4$-10.5 bin, where the signal disappears entirely when we remove filter B. This suggests that an individual bright source in the background is sufficient to raise the background enough to drown out the very weak signal from this bin.

\section{Null Tests}

Here we repeat the analysis in section 3, but instead of stacking locally brightest galaxies we stack random positions on the sky. Within each bin, we use the same number of random positions as the number of locally brightest galaxies in the bin, and we assign each random position the redshift of one of the locally brightest galaxies. The results are shown in Figure C1. The $1\sigma$ uncertainties are consistent with zero in 18/20 cases, the exceptions being the log $M_{\ast} = 10.3$-$10.4$ bin, which is a $3.0\sigma$ detection, and the log $M_{\ast} = 11.4$-$11.5$ bin, which is a $1.3\sigma$ detection.

\begin{figure}
\begin{center}
\includegraphics[width=8.5cm]{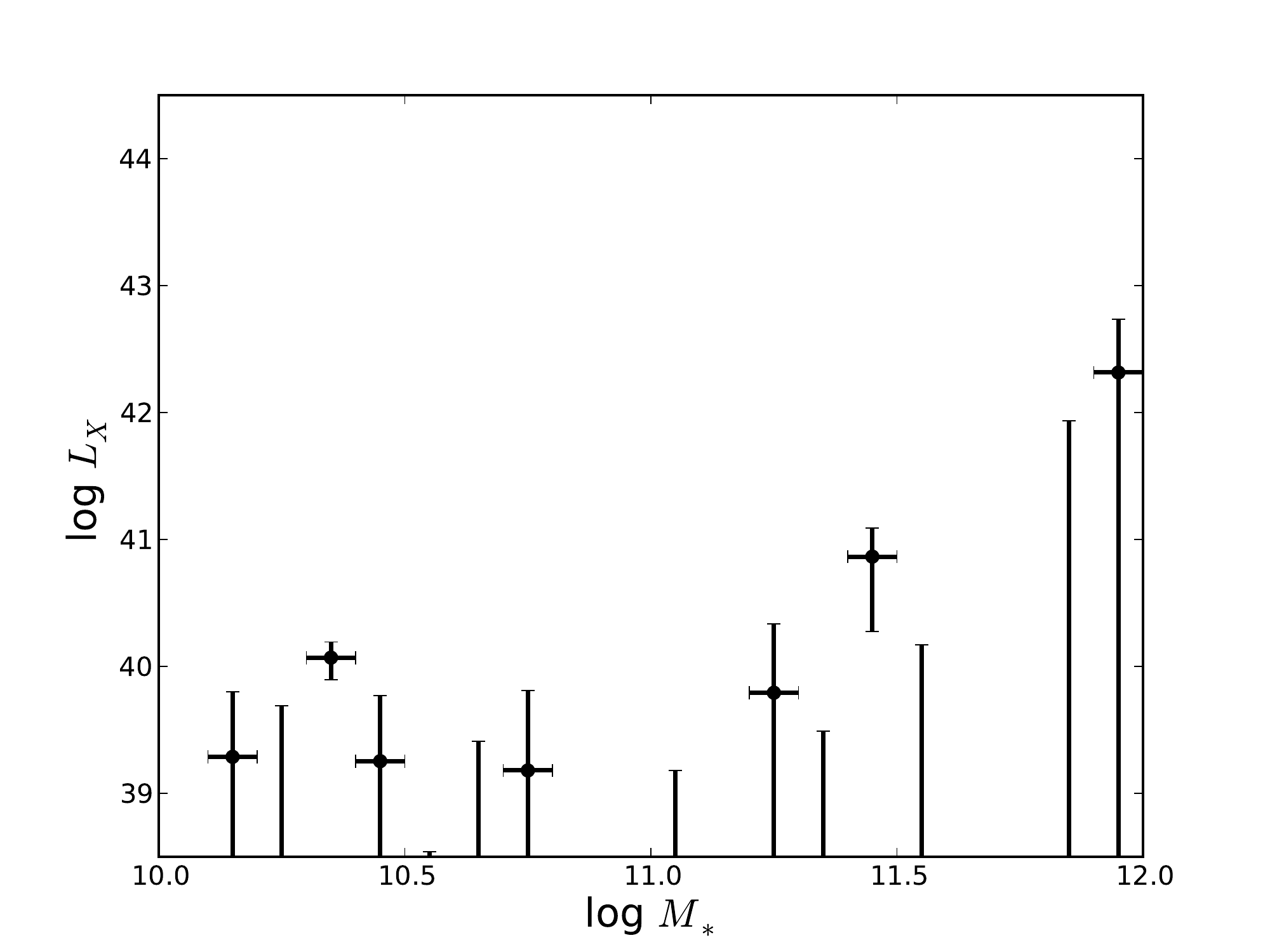}
\end{center}
\caption{Null tests showing the measured values of $L_{\text{total}}$ for stacks of random fields using the same assumed redshifts and numbers of galaxies as the locally brightest galaxies. Error bars show the measured $1\sigma$ uncertainties; the uncertainties are almost always consistent with zero, as expected.  }
\end{figure}

\section{Simulated Stacks}

 As described in Section 3, we have adapted our stacking analysis from ABD13 for use on locally brightest galaxies. In ABD13 we extensively tested our procedure on simulated data, but in this work we are studying more distant galaxies, and instead of modelling the image we are using the simpler technique of aperture photometry. In light of these changes, it is appropriate to verify that we can recover the correct $L_{500}$ with simulated data. 
 
 For each stellar mass bin, we draw galaxies from the simulated catalog (described in section 3), noting their stellar mass, halo mass, and the offset (in kpc) between the galaxy and the center of the halo. In most cases this offset is near zero, the primary failure mode turns out to be galaxies which lie more than 1 Mpc from the center of a galaxy cluster. When this occurs, the ''locally brightest galaxy'' is actually a satellite of a much more massive halo, but since the galaxy lies near the outskirts of the cluster this failure mode turns out not to be very serious. 

We draw the same number of galaxies from the simulated catalog as we have in the observed catalog, and we randomly assign each simulated galaxy the redshift of one of the observed galaxies. We assume an arbitrary $L_X$-$M_{500}$ relation and populate each simulated halo with X-ray luminosity distributed over a $\beta$-model \citep{Cavaliere1976} with $\beta = 0.6$. We assume the core radius is equal to the NFW scale radius, which we compute for each halo from its mass and redshift using the mass-concentration relation of \citet{Prada2012}. We then add a uniform background of $3\times10^{-4}$ count s$^{-1}$ arcmin$^{-2}$ to each simulated halo (with an assumed integration time of 400 s), stack them together in physical space, and convolve the resulting image with the empirical psf which we determined for that stellar mass bin (see section 3.1). We run our aperture photometry analysis on these simulated images, and compare the results to the input $L_X$. This comparison can be seen in Figure D1 for a variety of assumed $L_X$-$M_{500}$ relations which straddle the parameter space we search in the following section.

\begin{figure*}
\begin{center}
\includegraphics[width=17cm]{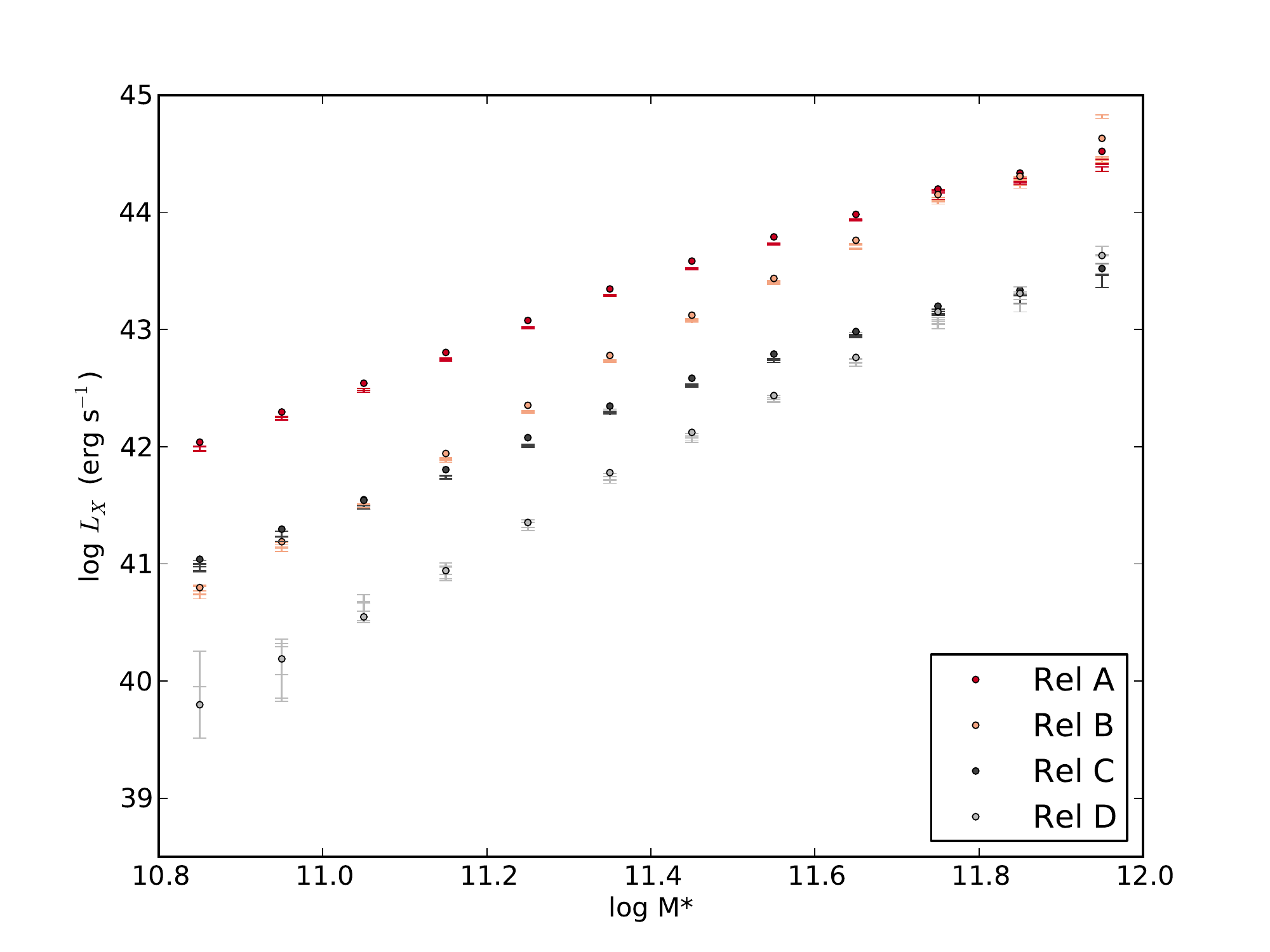}
\end{center}
\caption{Simulated emission from locally brightest galaxies, assuming four different power-law relations for the $L_X$-$M_{500}$ relation. For each relation, the data points represent the input relation using the effective $M_{\text{halo}}$ for each bin, and the error bars represent recovered values for $L_X$ after simulating realistic distributions of locally brightest galaxies, adding a background, stacking the galaxies, convolving with the psf, and performing aperture photometry. The recovered values match the input parameters very well, with deviations visible only for points which are several times fainter than the true values (Figure 5). The relations in this plot have the have same form as equation (6) and use the same values for $E(z)$, $C_{\text{bolo}}$, and $M_0$ as the true data. The values of ($L_0$, $\alpha$) for relations A, B, C, and D respectively are ($10^{45}$, $4/3$), ($10^{45}$, $2$), ($10^{44}$, $4/3$), and ($10^{44}$, $2$).  }
\end{figure*}

As Figure D1 shows, we can generally recover $L_X$ within the measurement errors across the entire range of $M_{\ast}$. The largest deviations occur at the high end where the intrinsic variation in $L_X$ is largest (due to the steep slope of the $M_{\ast}$-$M_{500}$ relation) at this end. There is also a very small systematic bias, but the magnitude of the bias is not significant and it seems most prominent for relation A, which has much higher normalization than our observed relation.

\section{X-ray Binary Contribution}

In order to estimate the effect of X-ray binaries on our measured X-ray luminosities, we make use of established scaling relations. LMXB emission is correlated with stellar mass (or equivalently with K-band stellar luminosity), so we can relate log $M_{\ast}$ to the expected LMXB signal. Converting the \citet{Boroson2011} relation to the 0.5-2.0 keV band, the expected LMXB luminosity is approximately

\begin{equation}
L_{\text{LMXB}} \approx 3\times10^{28} \frac{L_K}{L_{\odot}} \text{ erg s}^{-1} 
\end{equation}

For this figure we assume a very conservative K-band M/L ratio of 0.8 (the M/L ratio has a very weak dependence on stellar mass and on morphology, which we neglect) to convert this relation into a function of $M_{\ast}$. 

HMXB emission is correlated with star formation rate (SFR). To estimate the SFR, we use the ``B30'' indicator from the NYU Value-Added Galaxy Catalog (\citealt{Blanton2005}, \citealt{Blanton2007}) which estimates the fraction of the galaxy's stellar mass which has formed in the past 300 Myr. We multiply this by $M_{\ast}$ for each galaxy and divide by 300 Myr to get a SFR. Blanton and Roweis note that their B300 indicator underestimates the true star formation rate; in particular they show a 0.7 dex offset between results inferred using their indicator and the results of \citet{Hopkins2003}. We therefore correct the inferred SFR upwards by a factor of 0.7 dex to account for this offset. We then use the \citet{Mineo2012} relation (shifted to the 0.5-2.0 keV band):

\begin{equation}
L_{\text{HMXB}} \approx 1.4\times10^{39} \frac{\dot{M}}{1 M_{\odot} \text{ yr}^{-1}} \text{ erg s}^{-1} 
\end{equation}

We can get a very crude estimate of the expected HMXB signal by averaging together $L_{\text{HMXB}}$ for all the galaxies in each bin. The fraction of blue galaxies increases quickly as stellar mass decreases, so this actually leads to a prediction that $L_{\text{HMXB}}$ should increase towards less-massive galaxies in Figure 5.

We choose not to correct our results for XRB contamination, since this signal is negligible except in the log $M_{\ast} = 10.8-10.9$ and log $M_{\ast} = 10.9-11.0$bins, where XRBs contribute an estimated 38\% and 26\% of the emission, respectively. This is because the exact XRB contamination is highly uncertain: the scaling relations have intrinsic scatter and unknown bias, and the 0.7 dex correction to the star formation rate also contains considerable uncertainty. Moreover, these two bins also have the largest measurement uncertainties out of the twelve we consider, which means that they are given the least weight in the subsequent fits to the $L_X$-$M_{\ast}$ and $L_X$-$M_{500}$ relations, so we can expect that the XRB contamination is not affecting our results significantly. As a check, in the subsequent fits we repeated each analysis with an estimated XRB correction applied to these two bins, and the results did not change within the $1\sigma$ uncertainties.

\section{Cross-checking our $L_X$-$M_{500}$ relation results}

Since our results in section 5 rely on extensive forward-modeling of the  $L_X$-$M_{500}$  through a simulated catalog, it is appropriate to check how important the forward-modeling is for our results. To do this, we use the simpler technique of P13, and fit a linear relation to the $L_{\text{total}}$ results as a function of the effective $M_{500}$ in each bin. We fit relations of the form of equation 5, using the mean temperature (from Table 2) to compute the mean bolometric correction and the mean redshift (from the distance in Table 1) to compute the mean $E(z)$ in each bin. 

The resulting contours are shown in Figure E1, along with the contours from our forward-modeling for comparison. The best-fit linear relation yields a $\chi^2/$d.o.f. of $12.0/10$, which is unacceptable at $1\sigma$ but still allows for a $2\sigma$ and a $3\sigma$ contour region. The $2\sigma$ contour regions for both methods overlap nicely, although the forward-modeled relation prefers a slightly lower slope.

\begin{figure}
\begin{center}
\includegraphics[width=8.5cm]{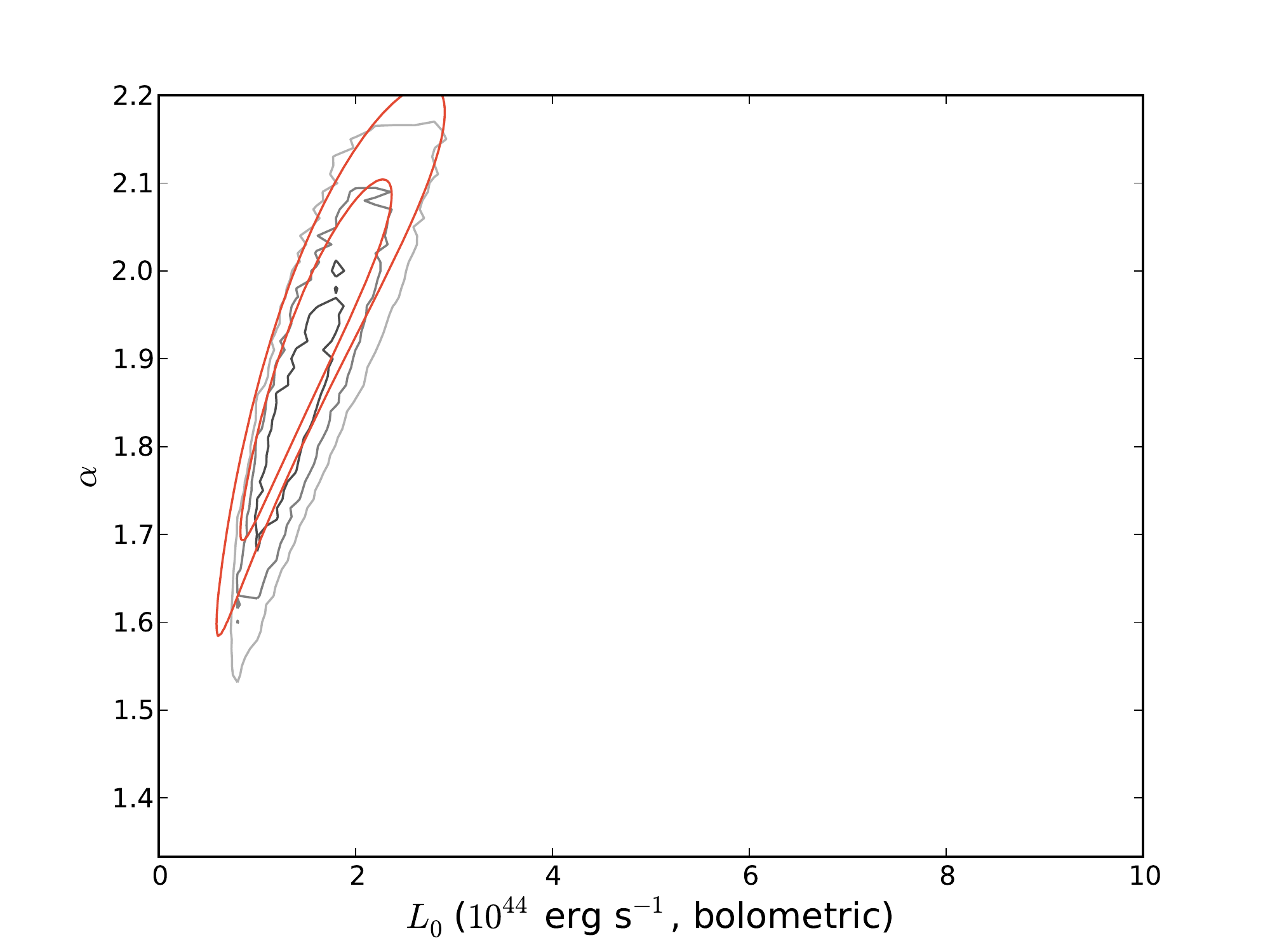}
\end{center}
\caption{Best-fit parameters for the $L_X$-$M_{500}$ relation, as measured by forward-modeling (gray contours) and by fitting to the effective $M_{500}$ values for each bin (red contour). The latter method gives less acceptable results (the best-fit relation has $\chi^2/$d.o.f. $=12.0/10$), so only the $2\sigma$ and $3\sigma$ contours can be plotted for this method. The $2\sigma$ contour regions for both methods overlap nicely, although the forward-modeled relation prefers a slightly lower slope. }
\end{figure}

\section{Total, Extrinsic, and Intrinsic Scatter in $L_X$}

In Table F1 we present the measured scatter in $L_{\text{total}}$ and $L_{\text{CGM}}$, and we distinguish between sources of extrinsic and intrinsic scatter. The ``measured'' scatter is estimated from the sample error on the mean as estimated from bootstrapping analysis, multiplied by the square root of the number of locally brightest galaxies in each bin. Since the mean number of photons per locally brightest galaxy is low (11.8 and 6.5 in our two highest-mass bins, decreasing down to 0.01 in the log $M_{\ast} = 10.1-10.2$ bin), we expect a significant contribution to this measured scatter from Poisson noise; this scatter is quantified in the ``Poisson'' columns. Another extrinsic source of scatter comes from the dispersion in the objects stacked within a given $M_{\ast}$ bin. This includes the dispersion in the $M_{500}$-$M_{\ast}$ relation, as well as the offsets between locally brightest galaxies and the centers of their dark matter halos. We estimate this scatter (which we denote as ``stacking'' scatter in Table F1) using our 10 different realizations of the simulated stacking analysis. We measure $L_{\text{total}}$ for each object in our simulated stacks, compute the standard deviation of $L_{\text{total}}$ in each stellar mass bin for each simulation, and take the mean of these ten realizations to estimate the stacking scatter. 

After accounting for these two forms of scatter, the remaining scatter (which we denote as ``other'' in Table F1) is due to scatter around the $L_X$-$M_{500}$ relation (intrinsic scatter) along with some unknown contribution from X-ray binaries and low-luminosity AGN activity, both of which likely become increasingly important at lower X-ray luminosities. The ``other'' scatter is therefore an upper limit on the amount of intrinsic scatter, although at the high-mass end we expect it to be close to the true intrinsic scatter. Towards the low-mass end the ``other'' scatter increases significantly. This is likely physical, reflecting increasing intrinsic variation in the X-ray properties of galaxies, and is probably compounded by the increasing contribution of blue galaxies to the sample as the stellar mass decreases.

\begin{table*}
\begin{minipage}{132mm}
\caption{ Total, Extrinsic, and Intrinsic Scatter }
\begin{tabular}{ccccccccc}
\hline
log $M_{\ast}$ & Total & Total & Total & Total & CGM & CGM & CGM & CGM\\
& Measured & Poisson & Stacking & Other & Measured & Poisson & Stacking & Other\\
\hline
11.9-12.0 & 1.31 & 0.26 & 0.72 & 1.06& 1.04 & 0.35 & 0.71 & 0.67\\
11.8-11.9 & 1.16 & 0.33 & 0.66 & 0.89& 1.16 & 0.41 & 0.66 & 0.86\\
11.7-11.8 & 1.63 & 0.34 & 0.73 & 1.41& 1.64 & 0.42 & 0.72 & 1.41\\
11.6-11.7 & 1.75 & 0.43 & 0.70 & 1.54& 1.52 & 0.52 &0.70 & 1.24\\
11.5-11.6 & 1.79 & 0.55 & 0.96 & 1.41& 2.07 & 0.64 & 0.99 & 1.70\\
11.4-11.5& 2.42 & 0.67 & 1.04 & 2.08& 2.46 & 0.83 & 1.08  & 2.05\\
11.3-11.4 & 2.70 & 0.87 & 1.02 & 2.34& 3.12 & 1.03 & 1.03 & 2.75\\
11.2-11.3 & 2.63 & 1.03 & 1.34 & 2.02& 3.17 & 1.23 & 1.40 &2.56\\
11.1-11.2 & 3.24 & 1.09 & 1.44 & 2.69& 3.67 & 1.32 & 1.49 & 3.08\\
11.0-11.1 & 3.76 & 1.35 & 1.78 & 3.03& 5.85 & 1.72 & 1.86 &5.28\\
10.9-11.0 & 3.70 & 1.59 & 1.95 & 2.72& 5.71 & 1.87  & 2.02 & 5.00\\
10.8-10.9 & 4.42 & 1.74 & 2.63 & 3.10& 7.02 & 2.89  &2.71 &5.80\\
10.7-10.8 & 5.59 & 2.17 & --  & 5.16& 4.93 & -- & -- & --\\
10.6-10.7 & 6.05 & 1.99 & -- & 5.71& 6.54 &-- &-- &--\\
10.5-10.6 & 6.38 & 2.46 & -- & 5.89& 6.22 &-- &-- &--\\
10.4-10.5 & 6.38 & 3.13 & -- &5.56 & 5.96 &-- &-- &--\\
10.3-10.4& 4.35 & 1.96 & -- & 3.89& 6.16 &2.37 &-- & 5.69\\
10.2-10.3 & 5.01 & 1.87 & -- &4.64 &6.15 &-- &-- & --\\
10.1-10.2 & 5.95 & 2.23  & -- &5.52 &5.92 &-- &-- &--\\
10.0-10.1 & 4.38 & -- & -- & --&4.38 &4.47 &-- &--\\
\hline
\end{tabular}
\\
\small{ Logarithmic scatter $\sigma_{\text{ln L}}$ per object as measured from our stacking analysis. See text for definitions of each column. Dashes in the stacking column refer to central galaxy stellar masses for which our assumed halo mass - X-ray luminosity relation is not an appropriate model for the observed emission. Dashes are indicated for the Poisson scatter in some bins because the net counts are negative.}
\end{minipage}
\end{table*}

\section{Imposing an X-ray Flux Limit}

In order to test our hypothesis that X-ray flux selection is responsible for the previous overestimates of the normalization of the $L_X$-$M_{500}$ relation, in this Appendix we impose an X-ray flux limit on our sample. We measure the flux from within $R_{500}$ for each galaxy, and only consider galaxies if they have a 0.5-2.0 keV observed-frame background-subtracted X-ray flux of at least $1\times10^{-12}$ erg s$^{-1}$ cm$^{-2}$ within $R_{500}$. This flux limit is lower than the limit used by REFLEX \citep{Bohringer2001}. However, we use a fixed aperture of size $R_{500}$ while they employed an adaptive aperture, which allows them to maximize the S/N of each cluster. Our flux limit still removes almost all of our sources: only 161 locally brightest galaxies remain, and above log $M_{\ast} = 11.9$ and log $M_{\ast} = 11.8$ there are only 2 and 8  objects respectively. Contamination from AGN and X-ray binaries also becomes more important, since we are now only considering a handful of objects.

In order to reduce the effects of this contamination, as well as to increase the number of data points available for fitting, we therefore eschew stacking entirely, and just fit a power-law relation with the form of equation 5 to the data. This requires assuming a conversion from stellar mass to halo mass. For the simple exercise in this Appendix, we use the conversion computed in Section 5 for the full LBG sample; the normalization of the relation cancels out so the slope is the major controlling parameter for this conversion and the slope we computed is in good agreement with other measurements (see Figure 6). We therefore neglect the effect that the assumed slope has on the inferred halo masses. We also restrict our attention to the galaxies with stellar masses above log $M_{\ast} = 11.6$, since these have temperatures above 2 keV and provide a more direct comparison to other studies (which have focused on galaxy clusters). Note that this also decreases the effect of the slope of the $L_X$-$M_{500}$ relation on our inferred masses, since it focuses on the systems which are closer to our pivot point. 

There are 47 locally brightest galaxies within this mass range that fall above our flux limit. The best-fit $L_X$-$M_{500}$ relation to these data has a $\chi^2$ of 5.2 for 45 degrees of freedom, which is very low. This reflects the large measurement errors on each galaxy (due to the fixed aperture size) as well as the propagated uncertainty due to the bolometric correction for each galaxy. The best-fit slope is $1.75_{-0.96}^{+0.93}$ and the best-fit normalization is $8.0_{-2.5}^{+2.3}\times10^{44}$ erg s$^{-1}$ ($1\sigma$ uncertainties). The uncertainty on the slope is large (due to the relatively narrow mass range), but if we fix the slope to be 1.84 (the value we measure in Section 5, which is fully consistent with the value measured by other studies), then this gives us self-consistent estimates of the halo masses, and the best-fit normalization is $8.2_{-1.4}^{+1.5} \times10^{44}$ erg s$^{-1}$. Both calculations clearly show that the normalization for the flux-limited sample is much higher than the inferred normalization for the full sample, and is much closer to the other X-ray flux-selected samples.

\end{document}